\documentclass[aps,prb,onecolumn,superscriptaddress,10pt]{revtex4-2}
\usepackage{amsmath,amssymb,bm,amsthm}
\usepackage{graphicx,float,xcolor}
\usepackage{orcidlink}
\usepackage[ruled]{algorithm2e}
\usepackage{hyperref}
\hypersetup{hidelinks}

\newcommand{\Tr}{\operatorname{Tr}}
\newcommand{\Log}{\operatorname{Log}}
\newcommand{\rank}{\operatorname{rank}}
\newcommand{\Span}{\operatorname{span}}
\newcommand{\EE}{\mathbb{E}}
\newcommand{\cH}{\mathcal{H}}
\newcommand{\cD}{\mathcal{D}}
\newcommand{\cN}{\mathcal{N}}
\newcommand{\cM}{\mathcal{M}}
\newcommand{\id}{\mathbb{I}}
\newcommand{\ket}[1]{\lvert #1\rangle}
\newcommand{\bra}[1]{\langle #1\rvert}
\newcommand{\braket}[2]{\langle #1\vert #2\rangle}
\newcommand{\abs}[1]{\lvert #1\rvert}
\newcommand{\norm}[1]{\lVert #1\rVert}

\newtheorem{proposition}{Proposition}

\begin{document}
\title{Entanglement and non-local magic in a non-unitarily deformed non-Hermitian bipartite system
}

		\author{Chen-Huan Wu 
			\orcidlink{0000-0003-1020-5977} }
		\thanks{chenhuanwu1@gmail.com}
        \affiliation{Department of Physics, Faculty of Science, Universiti Malaya, Kuala Lumpur 50603, Malaysia}

\begin{abstract}
\begingroup
Non-Hermitian degeneracies are usually discussed through spectral coalescence,
whereas entanglement is a property of eigenvectors and need not be fixed by the
eigenvalues alone.  We formulate a compact bipartite model that separates these
two notions.  A Hermitian operator with a degenerate eigenspace is transformed
by an invertible non-unitary similarity map.  The resulting Hamiltonian is
non-Hermitian and retains a non-defective degeneracy at every finite value of
the non-Hermiticity parameter.  For an exactly solvable two-qubit realization,
the canonical right eigenvectors obtained by transporting a fixed reference
basis evolve continuously from product states toward Bell states although the
spectrum is unchanged.  We distinguish the positive right-state reduced
density matrix from the generally non-positive biorthogonal reduction, for
which an eigenvalue-based entropy may become complex, and from complementary
SVD and bi-unit-invariant SVD diagnostics.  The same two-qubit solution gives
a closed partial-transpose negativity and an exact unrestricted second-order
non-local stabilizer R\'enyi entropy, whose computational-basis Schmidt
representative is globally optimal for the present rank-two spectrum.
Entanglement of the transported canonical states grows monotonically with
the deformation, whereas the non-local magic vanishes in both the product and maximally entangled
limits and is largest at an intermediate coupling.  By optimizing over
all normalized right eigenvectors in the degenerate doublet, we further obtain
an exact, basis-independent resource range: its minimum is zero, while its
maximum reaches $\log_2(4/3)$ at an intermediate deformation and remains
saturated thereafter, even as the canonical-state magic tends to zero.
In larger bipartite spaces, Page entropy and Haar-averaged purity provide reference values for
eigenstate typicality, subject to the symmetry sector and the actual tensor
product structure.  These diagnostics separate non-defective degeneracy,
exceptional-point sensitivity, Haar-typical entanglement, and non-stabilizer
correlations while distinguishing a chosen eigenbasis from the intrinsic
resource capacity of the full degenerate right eigenspace.
\endgroup
\end{abstract}


\maketitle

\section{Introduction}

\begingroup
Non-Hermitian Hamiltonians exhibit spectral and eigenvector structures that
have no direct counterpart in closed Hermitian systems
\cite{Ashida2020,Bergholtz2021}.  At an exceptional point, eigenvalues and
eigenvectors coalesce and the Hamiltonian becomes defective.  A repeated
eigenvalue, however, does not by itself imply an exceptional point: a
degenerate non-Hermitian Hamiltonian can remain diagonalizable when its
geometric and algebraic multiplicities agree \cite{Kato1995,Sayyad2022}.
This distinction is particularly important in bipartite and many-body
settings, where degeneracy, eigenvector nonorthogonality, and entanglement are
sometimes inferred from the same spectral information even though they are
logically independent properties.

An earlier random-matrix construction established that rank deficiency and
non-defective degeneracy can coexist in a non-Hermitian bipartite system
\cite{Wu2024}.  Here we formulate a more sharply controlled analytical
question: how can the bipartite resources carried by a diagonalizable
degenerate eigenspace change under an isospectral non-unitary deformation,
and how should those resources be compared with universal random-state
benchmarks?  We address this question using an invertible similarity
transformation that leaves every eigenvalue and its multiplicity unchanged
while continuously modifying the right eigenvectors.  This construction
separates the algebraic structure of the spectrum from the entanglement and
nonstabilizerness of the associated eigenstates.

The analysis requires a careful distinction among different reductions of a
non-Hermitian eigenstate.  A normalized right eigenvector produces an ordinary
positive density matrix with real von Neumann and R\'enyi entropies.  By
contrast, a biorthogonal dyad constructed from right and left eigenvectors has
unit trace but need not be Hermitian or positive; entropy expressions based on
its eigenvalues may consequently become negative or complex
\cite{Brody2014}.  Singular-value constructions provide complementary real
and non-negative diagnostics \cite{Caputa2024SVDPseudo,Caputa2026UISVD}.  In
the canonical two-qubit example considered below, the ordinary SVD entropy
reconstructs the positive right-state entropy, whereas the bi-unit-invariant
SVD entropy is invariant under admissible rescalings of the biorthogonal bases
and primarily detects the rank structure.  Only the positive right-state
reduction, defined with the fixed standard Hilbert-space inner product, has
the conventional interpretation required for a direct comparison with
Page's Haar-random result \cite{Page1993}.

Entanglement entropy alone does not determine how the Schmidt weights are
organized relative to stabilizer structure.  The non-local stabilizer
R\'enyi entropy removes local-basis dependence by minimizing the stabilizer
entropy over independent local unitaries.  Its computational-basis Schmidt
representative can be evaluated from Walsh--Hadamard autocorrelations of the
ordered Schmidt spectrum \cite{Torre2026}.  For the present state, whose
Schmidt rank is two, the global optimality theorem for Schmidt rank not
exceeding six guarantees that this representative realizes the unrestricted
local-unitary minimum \cite{Sierant2026SpectralGeometry}.  Thus
\begin{equation}
 \mathcal M_2^{\rm NL}(\psi_\gamma)
 =\mathcal M_2^{\rm Sch}(\psi_\gamma).
 \label{eq:IntroExactNonlocalMagic}
\end{equation}
For each transported canonical eigenvector, the optimized non-local magic
vanishes at the product and Bell endpoints and peaks at intermediate
deformation.  Maximization over the full degenerate eigenspace instead gives a
basis-independent resource envelope with a high-deformation plateau.  The
positive-partial-transpose criterion also supplies a complete separability
test for the positive two-qubit density matrix under mixing or dephasing
\cite{Peres1996,Baul2026}.

Section II introduces the isospectral deformation and its non-defective
eigenspace.  Section III gives the right-state and biorthogonal entanglement
and non-local magic results, including extrema over the degenerate doublet.
Sections IV and V establish the complex-Haar benchmarks and projected
subspace sampling for a general $\cN\times\cM$ bipartition, followed by the
numerical example in Sec. VI.  Section VII examines Pauli and replica
moments, Bell nonlocality, dephasing, dissipative preparation, magic
concentration, and symmetry-resolved information.

Although these operational representations are physically different, they do
not introduce independent state variables in the two-qubit model.  They are
fixed by the same pair
\begin{equation}
 u_\gamma=\operatorname{sech}(2|\gamma|),
 \qquad
 C_\gamma=\sqrt{1-u_\gamma^2}=\tanh(2|\gamma|),
 \label{eq:IntroResourceCoordinates}
\end{equation}
which respectively determine the imbalance and coherence of the Schmidt
spectrum.  For a selected canonical eigenvector these two coordinates fix the resource
geometry; the full degenerate right eigenspace additionally contains a
continuum of coherent superpositions with different Schmidt spectra.
Distinguishing the resources of a canonical vector from extrema over the
complete eigenspace is therefore essential to the interpretation of an
isospectral non-Hermitian deformation.  The resulting framework separates
non-defective degeneracy, exceptional-point sensitivity, Haar-typical
entanglement, and non-local nonstabilizerness without assigning physical
significance to an arbitrary basis choice inside a degenerate eigenspace.
\endgroup

\section{A diagonalizable non-Hermitian bipartite model}

Let
\begin{equation}
 \cH=\cH_A\otimes\cH_B,
 \qquad
 \dim\cH_A=\cN,
 \quad
 \dim\cH_B=\cM,
 \quad
 \cD=\cN\cM.
 \label{eq:hilbert}
\end{equation}
We start from a Hermitian reference operator with a degenerate sector,
\begin{equation}
 H_0=\lambda_0\Pi_0+\sum_{a=1}^{q}\lambda_a\Pi_a,
 \qquad
 \Pi_a\Pi_b=\delta_{ab}\Pi_a,
 \qquad
 \sum_{a=0}^{q}\Pi_a=\id,
 \label{eq:H0}
\end{equation}
where $\Pi_0 = \sum_{\mu=1}^r \vert{}0,\mu\rangle \langle 0,\mu\vert{} $ and $\rank\Pi_0=r>1$ and the displayed eigenvalues are distinct.  For an
invertible, generally non-unitary operator $S_\gamma$, define
\begin{equation}
 H_\gamma=S_\gamma H_0S_\gamma^{-1}.
 \label{eq:Hsimilar}
\end{equation}
The parameter $\gamma$ controls non-Hermiticity.  A useful family is
\begin{equation}
 S_\gamma=e^{\gamma G},
 \qquad G=G^\dagger,
 \label{eq:Sgeneral}
\end{equation}
where the non-local Hermitian operator $G= \sigma_x \otimes \sigma_x$ (with eigenvalues $\pm 1$) containing a genuine $A$--$B$ coupling and $S_\gamma^\dagger 
= e^{\gamma G^\dagger} = e^{\gamma G} = S_\gamma$. The generator $G = \sigma_x \otimes \sigma_x$ explicitly breaks the continuous global $U(1)$ symmetry of the system by coupling the zero-excitation state $\vert{}00\rangle$ with the two-excitation state $\vert{}11\rangle$.
\begingroup
Since $G=G^\dagger$, $S_\gamma=e^{\gamma G}$ is positive definite for real
$\gamma$.  Its singular values coincide individually with its positive
eigenvalues, which in the present case are $e^\gamma$ and $e^{-\gamma}$.
Moreover, the exponential map is invertible for every finite real $\gamma$,
with $S_\gamma = e^{\gamma G} = \cosh\gamma \, \mathbb{I} + \sinh\gamma \, G$, $S_\gamma^{-1} = e^{-\gamma G} = \cosh\gamma \, \mathbb{I} - \sinh\gamma \, G$
; hence invertibility is automatic along the
entire deformation.
\endgroup

If $\ket{a,\mu}$ is an orthonormal eigenbasis of $H_0$, the right and left
eigenvectors of $H_\gamma$ can be chosen as
\begin{equation}
 \ket{R_{a\mu}}=S_\gamma\ket{a,\mu},
 \qquad
 \bra{L_{a\mu}}=\bra{a,\mu}S_\gamma^{-1}.
 \label{eq:LR}
\end{equation}
They obey
\begin{equation}
\begin{aligned}
    &
 H_\gamma \vert{}R_{a\mu}\rangle = (S_\gamma H_0 S_\gamma^{-1}) (S_\gamma \vert{}a,\mu\rangle) = S_\gamma H_0 \vert{}a,\mu\rangle = \lambda_a (S_\gamma \vert{}a,\mu\rangle) = \lambda_a \vert{}R_{a\mu}\rangle,\\
&\bra{L_{a\mu}}H_\gamma=\lambda_a\bra{L_{a\mu}},
 \qquad
 \braket{L_{a\mu}}{R_{b\nu}}=\delta_{ab}\delta_{\mu\nu}.
 \label{eq:biorth}
\end{aligned}
\end{equation}
\begingroup
The standard invariance of Jordan structure under similarity
\cite{Kato1995} fixes the multiplicity of the deformed eigenspace.
\begin{proposition}
Let $H_0=\sum_a\lambda_a\Pi_a$ be Hermitian, with
$\operatorname{rank}\Pi_0=r$, and let
$H_\gamma=S_\gamma H_0S_\gamma^{-1}$ for an invertible operator
$S_\gamma$.  Then $H_\gamma$ is diagonalizable and has the same eigenvalues,
algebraic multiplicities, geometric multiplicities, and Jordan-block
structure as $H_0$.  In particular,
\begin{equation}
 \dim\ker(H_\gamma-\lambda_0\id)
 =\dim\ker(H_0-\lambda_0\id)=r,
 \label{eq:KernelMultiplicityPreservation}
\end{equation}
so the multiplicity-$r$ eigenvalue $\lambda_0$ remains non-defective.
\end{proposition}

\begin{proof}
Since
$H_\gamma-\lambda\id=S_\gamma(H_0-\lambda\id)S_\gamma^{-1}$,
every $|v\rangle\in\ker(H_0-\lambda\id)$ is mapped to
$S_\gamma|v\rangle\in\ker(H_\gamma-\lambda\id)$.  Conversely, applying
$S_\gamma^{-1}$ maps every vector in the latter kernel back to the former.
Hence
\begin{equation}
 \ker(H_\gamma-\lambda\id)
 =S_\gamma\ker(H_0-\lambda\id),
 \label{eq:KernelSimilarityMap}
\end{equation}
and invertibility of $S_\gamma$ preserves the dimension of every eigenspace.
Similarity also preserves the characteristic and minimal polynomials and
therefore the full Jordan canonical form.  Because the Hermitian operator
$H_0$ is diagonalizable, $H_\gamma$ is diagonalizable as well.
\end{proof}

For $S_\gamma=e^{\gamma G}$, the inverse $e^{-\gamma G}$ exists for every
finite real $\gamma$.  The degenerate sector therefore remains
non-defective throughout the finite-parameter deformation.

The associated spectral projector onto the degenerate subspace is the oblique
projector
\begin{align}
 P_0^{RL}
 &=S_\gamma\Pi_0S_\gamma^{-1}
 =S_\gamma\left(\sum_{\mu=1}^r|0,\mu\rangle\langle0,\mu|\right)
 S_\gamma^{-1}\notag\\
 &=\sum_{\mu=1}^r(S_\gamma|0,\mu\rangle)
 (\langle0,\mu|S_\gamma^{-1})
 =\sum_{\mu=1}^{r}\ket{R_{0\mu}}\bra{L_{0\mu}},
 \qquad (P_0^{RL})^2=P_0^{RL}.
 \label{eq:oblique}
\end{align}
which satisfies the following operator properties:
\begin{itemize}
\item \textbf{Idempotence:} $(P_0^{RL})^2 = (S_\gamma \Pi_0 S_\gamma^{-1})(S_\gamma \Pi_0 S_\gamma^{-1}) = S_\gamma \Pi_0 (S_\gamma^{-1} S_\gamma) \Pi_0 S_\gamma^{-1} = S_\gamma \Pi_0^2 S_\gamma^{-1} = P_0^{RL}$.
\item \textbf{Oblique (Non-Hermitian) character:} For $\gamma \ne 0$, $(P_0^{RL})^\dagger = S_\gamma^{-1} \Pi_0 S_\gamma \ne P_0^{RL}$ because $S_\gamma^\dagger = S_\gamma \ne S_\gamma^{-1}$. Its range and kernel are complementary but not mutually orthogonal, representing a canonical biorthogonal oblique projection.
    \item \textbf{Subspace basis independence:} While individual degenerate right eigenvectors $|R_{0\mu}\rangle$ depend explicitly on the selected reference basis inside the degenerate sector, $\Pi_0$ is the unique, basis-invariant orthogonal projector onto $\ker(H_0 - \lambda_0 \mathbb{I})$ (invariant under any internal unitary rotation $U \Pi_0 U^\dagger = \Pi_0$). Consequently, the global oblique projector $P_0^{RL}$ is an operator invariant of the non-Hermitian eigenspace as a whole.
\end{itemize}
This basis independence is crucial: an entanglement entropy assigned to a single eigenvector inside a degenerate eigenspace is basis dependent and non-unique, whereas an ensemble sampled uniformly from the complete eigenspace via $Q_0^R$ defines an invariant physical benchmark.
$P_0^{RL}$ is basis independent inside the degenerate sector (due to the invariance $U \Pi_0 U^\dagger = \Pi_0$), even though individual
eigenvectors are not (like $\vert{}R_{0\mu}\rangle$ and $\langle L_{0\mu}\vert{}$).  

For the exponential deformation used below, no exceptional point can occur at
finite $\gamma$ because the similarity map never loses invertibility.  Since
$S_\gamma$ is Hermitian and positive, the Hamiltonian is also
quasi-Hermitian with metric
\begin{equation}
 \eta_\gamma=(S_\gamma^{-1})^\dagger S_\gamma^{-1}
 =S_\gamma^{-2},
 \qquad
 H_\gamma^\dagger\eta_\gamma=\eta_\gamma H_\gamma.
 \label{eq:QuasiHermitianMetric}
\end{equation}
We evaluate the normalized right states using the standard Hilbert-space
inner product and a fixed $A|B$ tensor-product decomposition.  For an
effective non-Hermitian description with fixed laboratory observables, as in
conditional or postselected dynamics, the nonlocal map $S_\gamma$ changes the
right-state entanglement while preserving the spectrum.  A quasi-Hermitian
description instead transforms the observables together with the metric
$\eta_\gamma=S_\gamma^{-2}$; the similarity map then represents a change of
description.  Since $S_\gamma$ acts nonlocally across $A|B$, this change also
affects the representation of subsystem-local observables.  The resource
analysis below uses the fixed-inner-product, fixed-bipartition convention.
\endgroup

For a reference eigenvector $|n\rangle$, introduce the norm
$N_n(\gamma)=\sqrt{\langle n|e^{2\gamma G}|n\rangle}$ and the normalized
right state $|\widetilde R_n\rangle=N_n^{-1}e^{\gamma G}|n\rangle$.
Because $G=G^\dagger$, differentiation gives
$\partial_\gamma N_n=N_n\langle G\rangle_n$ and hence
\begin{align*}
 \partial_\gamma|\widetilde R_n\rangle
 &=(G-\langle G\rangle_n)|\widetilde R_n\rangle,\\
 \langle\widetilde R_n|\partial_\gamma\widetilde R_n\rangle&=0.
\end{align*}
The corresponding Fubini--Study susceptibility is therefore
\begin{equation*}
 \chi_\gamma^R
 =\langle\partial_\gamma\widetilde R_n|
   \partial_\gamma\widetilde R_n\rangle
 -|\langle\widetilde R_n|\partial_\gamma\widetilde R_n\rangle|^2
 =\langle G^2\rangle_n-\langle G\rangle_n^2.
\end{equation*}
For the two-qubit generator $G=\sigma_x\otimes\sigma_x$, one has
$G^2=I_4$.  In particular, the normalized state transported from
$|00\rangle$ obeys $\langle G\rangle_{00}=\tanh(2\gamma)$, which gives
$\chi_\gamma^R=\operatorname{sech}^2(2\gamma)$.

Because $G^2=I_4$, the similarity map reduces to
$e^{\gamma G}=\cosh\gamma\,I_4+\sinh\gamma\,G$.  Thus
\begin{equation*}
 |R_{00}\rangle=\cosh\gamma|00\rangle+\sinh\gamma|11\rangle,
 \qquad
 |\widetilde R_{00}\rangle
 =\frac{\cosh\gamma|00\rangle+\sinh\gamma|11\rangle}
 {\sqrt{\cosh(2\gamma)}}.
\end{equation*}

\section{Right-state and biorthogonal entanglement}

\begingroup
Throughout this work, the bipartition and the standard Hilbert-space inner
product are held fixed while the right eigenvectors are deformed.  This
convention must be distinguished from a metric-redefined quasi-Hermitian
description in which observables and the inner product are transformed
together with the Hamiltonian.  Accordingly, for a right eigenvector we use
the positive right-state reduction defined with respect to the fixed standard
inner product and the fixed tensor-product decomposition
$\cH_A\otimes\cH_B$:
\endgroup
\begin{equation}
 \rho_A^{R}=\Tr_B\left(
 \frac{\ket{R}\bra{R}}{\braket{R}{R}}
 \right).
 \label{eq:rightdensity}
\end{equation}
Its von Neumann and second R\'enyi entropies are
\begin{align}
 S_A^{(\mathrm{vN},R)}&=-\Tr\rho_A^R\ln\rho_A^R,
 \label{eq:SvNright}\\
 S_A^{(2,R)}&=-\ln\Tr[(\rho_A^R)^2].
 \label{eq:S2right}
\end{align}
Both are real and satisfy
\begin{equation}
 0\le S_A^{(2,R)}\le S_A^{(\mathrm{vN},R)}
 \le\ln\min(\cN,\cM).
 \label{eq:bounds}
\end{equation}

The biorthogonal alternative is
\begin{equation}
 \rho_A^{RL}=\Tr_B\left(
 \frac{\ket{R}\bra{L}}{\braket{L}{R}}
 \right).
 \label{eq:biodensity}
\end{equation}
Although $\Tr\rho_A^{RL}=1$, this operator is generally non-Hermitian.  The
formal quantities
\begin{equation}
 S_A^{(\mathrm{vN},RL)}=-\Tr(\rho_A^{RL}\Log\rho_A^{RL}),
 \qquad
 S_A^{(2,RL)}=-\Log\Tr[(\rho_A^{RL})^2]
 \label{eq:bioentropies}
\end{equation}
depend on a branch of the matrix logarithm and should not be interpreted as
ordinary entanglement entropies.  We retain them only as diagnostics of
biorthogonal non-positivity.

\subsection{Exactly solvable two-qubit realization}

Take $\cN=\cM=2$ and use the ordered basis
$\{\ket{00},\ket{01},\ket{10},\ket{11}\}$
(we denote $\ket{00}=|0\rangle_A \otimes |0\rangle_B$) in Hilbert space $\mathbb{C}^2 \otimes \mathbb{C}^2$.  Let
\begin{equation}
 H_0=\operatorname{diag}(\lambda_0,\lambda_0,\lambda_1,\lambda_2),
 \qquad
 S_\gamma=e^{\gamma\sigma_x\otimes\sigma_x}
 =c\,\id+s\,\sigma_x\otimes\sigma_x,
 \label{eq:twoqubit}
\end{equation}
where $c=\cosh\gamma$ and $s=\sinh\gamma$. Thus $H_0 = \lambda_0 \Big(\vert{}00\rangle\langle 00\vert{} + \vert{}01\rangle\langle 01\vert{}\Big) + \lambda_1 \vert{}10\rangle\langle 10\vert{} + \lambda_2 \vert{}11\rangle\langle 11\vert{}$ with $\Pi_0 = \sum_{\mu=1}^2 \vert{}0,\mu\rangle\langle 0,\mu\vert{}=\vert{}00\rangle\langle 00\vert{} + \vert{}01\rangle\langle 01\vert{}$. Direct multiplication gives
\begin{equation}
H_\gamma=
\begin{pmatrix}
c^2\lambda_0-s^2\lambda_2&0&0&cs(\lambda_2-\lambda_0)\\
0&c^2\lambda_0-s^2\lambda_1&cs(\lambda_1-\lambda_0)&0\\
0&cs(\lambda_0-\lambda_1)&c^2\lambda_1-s^2\lambda_0&0\\
cs(\lambda_0-\lambda_2)&0&0&c^2\lambda_2-s^2\lambda_0
\end{pmatrix}.
\label{eq:Hmatrix}
\end{equation}
For real nonzero $\gamma$ and distinct $\lambda_a$, this matrix is real but
nonsymmetric, hence non-Hermitian.  Its spectrum remains
$\{\lambda_0,\lambda_0,\lambda_1,\lambda_2\}$.

The two right eigenvectors at the degenerate eigenvalue are
\begin{align}
 \ket{R_{00}}&=c\ket{00}+s\ket{11},\\
 \ket{R_{01}}&=c\ket{01}+s\ket{10}.
 \label{eq:Rdeg}
\end{align}
They are linearly independent and mutually orthogonal, with the common norm
$Z_\gamma=c^2+s^2=\cosh(2\gamma)$.
Either normalized right state has Schmidt probabilities
\begin{equation}
 p_\pm=\frac12\left[1\pm\operatorname{sech}(2\gamma)\right].
 \label{eq:schmidt}
\end{equation}
Consequently,
\begin{align}
 S_A^{(\mathrm{vN},R)}(\gamma)
 &=-p_+\ln p_+-p_-\ln p_- ,
 \label{eq:exactSvN}\\
 S_A^{(2,R)}(\gamma)
 &= -\ln\operatorname{Tr}(\rho_A^2) =-ln(p_+^2+p_-^2)=-\ln\left[1-\frac12\tanh^2(2\gamma)\right].
 \label{eq:exactS2}
\end{align}
At $\gamma=0$, $p_+ = 1, p_- = 0$, the states are separable.  For $\abs{\gamma}\to\infty$,
$p_\pm\to1/2$ and both entropies approach $\ln2$, reflecting the maximally entangled state.  Thus an unchanged,
\begingroup
non-defectively degenerate spectrum is compatible, within the canonical basis
obtained by transporting the fixed reference basis through $S_\gamma$, with a
continuous interpolation from product to maximally entangled right
eigenvectors.  This statement concerns the selected transported eigenbasis;
individual entanglement values are not invariants of the degenerate
eigenspace.
\endgroup

The corresponding left vector is
$\bra{L_{00}}=c\bra{00}-s\bra{11}$, with
$\braket{L_{00}}{R_{00}}=1$.
Its biorthogonal reduction is
\begin{equation}
 \rho_A^{RL}=c^2\ket{0}\bra{0}-s^2\ket{1}\bra{1}.
 \label{eq:biotwo}
\end{equation}
The negative eigenvalue immediately shows why Eq.~\eqref{eq:bioentropies}
cannot be compared with a Page value.  On the principal branch,
\begin{align}
 S_A^{(\mathrm{vN},RL)}
 &=-c^2\ln c^2+s^2\ln s^2+i\pi s^2,
 \label{eq:complexS}\\
 S_A^{(2,RL)}&=-\ln(c^4+s^4).
\end{align}

\begingroup

The same reduced transition operator admits a complementary singular-value
description.  The ordinary SVD entropy is real and bounded because it is
constructed from the normalized singular values of the transition operator
\cite{Caputa2024SVDPseudo,Caputa2026UISVD}.  For
Eq.~\eqref{eq:biotwo}, these singular values are
$\sigma_1=c^2$ and $\sigma_2=s^2$, and hence
\begin{equation}
 \frac{\sigma_1}{\sigma_1+\sigma_2}
 =\frac{c^2}{c^2+s^2}=p_+,
 \qquad
 \frac{\sigma_2}{\sigma_1+\sigma_2}
 =\frac{s^2}{c^2+s^2}=p_-.
 \label{eq:BiorthogonalNormalizedSingularValues}
\end{equation}
It follows immediately that
\begin{equation}
 S_A^{(\mathrm{SVD},RL)}
 =-p_+\ln p_+-p_-\ln p_-
 =S_A^{(\mathrm{vN},R)}.
 \label{eq:SVDEntropyEqualsRightEntropy}
\end{equation}
Thus the ordinary SVD entropy of the canonical biorthogonal reduction
reconstructs the positive right-state entanglement entropy in the fixed
standard inner product, whereas the
eigenvalue-based quantity in Eq.~\eqref{eq:complexS} retains the sign and
branch information associated with non-positivity.

The bi-unit-invariant SVD construction removes arbitrary nonzero diagonal
units from both sides of a transition matrix before evaluating its singular
values \cite{Caputa2026UISVD}.  For
$\gamma\neq0$, one admissible balancing of Eq.~\eqref{eq:biotwo} is
\begin{equation}
 D_{BL}=D_{BR}
 =\operatorname{diag}\!\left(c^{-1},|s|^{-1}\right),
 \qquad
 D_{BL}\rho_A^{RL}D_{BR}
 =\operatorname{diag}(1,-1).
 \label{eq:BiorthogonalBUIBalancing}
\end{equation}
The two BUI singular values are therefore equal and the corresponding entropy
is
\begin{equation}
 S_A^{(\mathrm{BUI},RL)}
 =
 \begin{cases}
  0,&\gamma=0,\\
  \ln2,&\gamma\neq0.
 \end{cases}
 \label{eq:BiorthogonalBUISVDEntropy}
\end{equation}
The jump at $\gamma=0$ occurs because the rank changes and the balancing
factor $|s|^{-1}$ becomes singular.  Consequently, this BUI entropy is real,
non-negative, bounded, and invariant under the allowed diagonal rescalings,
but in the present diagonal model it is essentially a rank-sensitive
diagnostic rather than a continuous measure of the amount of physical
entanglement.  The eigenvalue-based RL entropy is already invariant under the
admissible diagonal similarity transformation, while ordinary SVD entropy is
not invariant under independent left and right diagonal rescalings in a
generic non-diagonal problem.  For the particular diagonal operator in
Eq.~\eqref{eq:biotwo}, however, the admissible biorthogonal similarity
rescalings leave even its ordinary singular values unchanged.

\endgroup

The condition number of the similarity map is
\begin{equation}
 \kappa_2(S_\gamma)= \frac{e^{\vert{}\gamma\vert{}}}{e^{-\vert{}\gamma\vert{}}} = \frac{\sigma_{\max}(S_\gamma)}{\sigma_{\min}(S_\gamma)}=e^{2\abs{\gamma}}.
 \label{eq:kappa}
\end{equation}
Since $G = G^\dagger$, the positive singular values of $S_\gamma = e^{\gamma G}$ coincide identically with its positive eigenvalues, $\{e^{|\gamma|}, e^{-|\gamma|}\}$. In non-Hermitian spectral theory, an exceptional point (EP) requires the algebraic multiplicity of a degenerate eigenvalue to exceed its geometric multiplicity, forcing the coalescing eigenvectors to span an incomplete basis associated with a non-trivial Jordan block of dimension $\ge 2$. Under an isospectral similarity mapping from a diagonalizable reference operator $H_0$, an EP can occur only if the similarity map $S_\gamma$ becomes singular, or equivalently, if its condition number diverges, $\kappa_2(S_\gamma) \to \infty$. In the present model, because $S_\gamma = e^{\gamma G}$ has full operator rank and bounded generator $\|G\|_\infty = 1$, the condition number $\kappa_2(S_\gamma) = e^{2|\gamma|}$ remains finite for every finite real parameter $\gamma \in (-\infty, \infty)$. Thus the system is algebraically protected against eigenvector coalescence, ensuring that the multiplicity-$r$ degeneracy remains non-defective across the entire finite parameter domain, while the asymptotic divergence $\kappa_2 \to \infty$ as $|\gamma| \to \infty$ reflects entanglement saturation rather than a finite-coupling spectral singularity.
Entanglement saturation and spectral
sensitivity can therefore grow together without an exceptional point at any
finite parameter value.

\subsection{Schmidt reduction and participation statistics}

We now derive Eq.~\eqref{eq:schmidt} explicitly and clarify why the
computational-basis inverse participation ratio happens to coincide with the
reduced purity in this particular model.  The normalized first right
eigenvector in Eq.~\eqref{eq:Rdeg} is
\begin{equation}
 \ket{\widetilde R_{00}}
 =\frac{c\ket{00}+s\ket{11}}{\sqrt{Z_\gamma}}
 =a\ket{0}_A\ket{0}_B+b\ket{1}_A\ket{1}_B,
 \qquad
 a=\frac{c}{\sqrt{Z_\gamma}},\quad
 b=\frac{s}{\sqrt{Z_\gamma}}.
 \label{eq:R00normalized}
\end{equation}
The two-dimensional basis $\{\ket{0}_A,\ket{1}_A\}$ and
$\{\ket{0}_B,\ket{1}_B\}$ are orthonormal, and the coefficients are real and non-negative.  Equation
\eqref{eq:R00normalized} is therefore already a Schmidt decomposition, up to
an irrelevant sign of $b$ when $\gamma<0$, which can be absorbed into either
local Schmidt vector.  In particular, no nontrivial local basis rotation is
needed: the computational product basis is aligned with the Schmidt basis.

To display the partial trace step by step, expand the pure-state density
operator as
\begin{equation}
\begin{aligned}
\rho&= \ket{\widetilde R_{00}}\bra{\widetilde R_{00}}
 =\frac{1}{Z_\gamma}\bigl(c^2\ket{00}\bra{00}
 +cs\ket{00}\bra{11}
 +cs\ket{11}\bra{00}
 +s^2\ket{11}\bra{11}\bigr)\\
& = a^2 \vert{}00\rangle\langle00\vert{} + ab \vert{}00\rangle\langle11\vert{} + ab \vert{}11\rangle\langle00\vert{} + b^2 \vert{}11\rangle\langle11\vert{}.
 \label{eq:R00densityexpanded}
\end{aligned}
\end{equation}
For product-basis operators, the partial trace satisfies
\begin{equation}
 \Tr_B\!\left(\ket{i}_A\ket{j}_B
 \bra{k}_A\bra{l}_B\right)
 =\braket{l}{j}\ket{i}_A\bra{k}_A
 =\delta_{jl}\ket{i}_A\bra{k}_A.
 \label{eq:partialtraceidentity}
\end{equation}
Consequently, the two cross terms in
Eq.~\eqref{eq:R00densityexpanded} vanish because
$\braket{1}{0}=\braket{0}{1}=0$, while the diagonal terms survive.  This gives
\begin{align}
 \rho_A^R
 &=\Tr_B\left(\ket{\widetilde R_{00}}
 \bra{\widetilde R_{00}}\right) \notag\\
& =\langle 0_B \vert{}\widetilde{R}_{00}\rangle\langle\widetilde{R}_{00}\vert{} 0_B \rangle + \langle 1_B \vert{}\widetilde{R}_{00}\rangle\langle\widetilde{R}_{00}\vert{} 1_B \rangle\\
 &=\frac{c^2}{Z_\gamma}\ket{0}_{A}\bra{0}_{A}
 +\frac{s^2}{Z_\gamma}\ket{1}_{A}\bra{1}_{A},
 \label{eq:rhoAexplicit}
\end{align}
where $\langle 0_B \vert{} 00\rangle = \vert{}0_A\rangle$ and $\langle 1_B \vert{} 11\rangle = \vert{}1_A\rangle$.
Its eigenvalues are therefore
\begin{equation}
 p_+=a^2=\frac{\cosh^2\gamma}{\cosh(2\gamma)},
 \qquad
 p_-=b^2=\frac{\sinh^2\gamma}{\cosh(2\gamma)}.
 \label{eq:psquaredforms}
\end{equation}
Using $\cosh^2\gamma=[\cosh(2\gamma)+1]/2$,
$\sinh^2\gamma=[\cosh(2\gamma)-1]/2$, and
$\cosh^2\gamma+\sinh^2\gamma=\cosh(2\gamma)$, one immediately obtains
$p_\pm=\{1\pm\operatorname{sech}(2\gamma)\}/2$ and $p_++p_-=1$,
which proves Eq.~\eqref{eq:schmidt}.  The normalized state derived from
$\ket{R_{01}}$ has the form $a\ket{0}_A\ket{1}_B+
b\ket{1}_A\ket{0}_B$ and hence has exactly the same pair of Schmidt
probabilities.
For eigenvalue $\lambda_1$, $\vert{}R_{10}\rangle = S_\gamma \vert{}10\rangle = (c\mathbb{I} + s \sigma_x \otimes \sigma_x)\vert{}10\rangle = c\vert{}10\rangle + s\vert{}01\rangle$,
$\vert{}\tilde{R}_{10}\rangle = \frac{c\vert{}10\rangle + s\vert{}01\rangle}{\sqrt{\cosh(2\gamma)}} = a\vert{}1\rangle_A\vert{}0\rangle_B + b\vert{}0\rangle_A\vert{}1\rangle_B$, and $\operatorname{Tr}_B(\vert{}\tilde{R}_{10}\rangle\langle\tilde{R}_{10}\vert{}) = a^2\vert{}1\rangle\langle 1\vert{}_A + b^2\vert{}0\rangle\langle 0\vert{}_A = p_+\vert{}1\rangle\langle 1\vert{}_A + p_-\vert{}0\rangle\langle 0\vert{}_A$.
For eigenvalue $\lambda_2$, $\vert{}R_{11}\rangle = S_\gamma \vert{}11\rangle = (c\mathbb{I} + s \sigma_x \otimes \sigma_x)\vert{}11\rangle = c\vert{}11\rangle + s\vert{}00\rangle = s\vert{}00\rangle + c\vert{}11\rangle$,
$\vert{}\tilde{R}_{11}\rangle = \frac{s\vert{}00\rangle + c\vert{}11\rangle}{\sqrt{\cosh(2\gamma)}} = b\vert{}0\rangle_A\vert{}0\rangle_B + a\vert{}1\rangle_A\vert{}1\rangle_B$, $\operatorname{Tr}_B(\vert{}\tilde{R}_{11}\rangle\langle\tilde{R}_{11}\vert{}) = b^2\vert{}0\rangle\langle 0\vert{}_A + a^2\vert{}1\rangle\langle 1\vert{}_A = p_-\vert{}0\rangle\langle 0\vert{}_A + p_+\vert{}1\rangle\langle 1\vert{}_A$.

Similarly, $\vert{}\tilde{R}_{01}\rangle = a\vert{}01\rangle + b\vert{}10\rangle = a\vert{}0\rangle_A\vert{}1\rangle_B + b\vert{}1\rangle_A\vert{}0\rangle_B$,
and  $\operatorname{Tr}_B(\vert{}\tilde{R}_{01}\rangle\langle\tilde{R}_{01}\vert{}) = a^2 \vert{}0\rangle\langle0\vert{} + b^2 \vert{}1\rangle\langle1\vert{} = p_+ \vert{}0\rangle\langle0\vert{} + p_- \vert{}1\rangle\langle1\vert{}$.

Let $q_x=\abs{\braket{x}{\widetilde R_{00}}}^2$ denote the probability of a
global computational-basis outcome $x\in\{00,01,10,11\}$. $q_x^2 = \Big[ \text{Tr}(\vert{}x\rangle\langle x\vert{} \Psi) \Big]^2 = \text{Tr}\Big[ (\vert{}x\rangle\langle x\vert{})^{\otimes 2} \Psi^{\otimes 2} \Big]$. In the present
case, $(q_{00},q_{01},q_{10},q_{11})=(p_+,0,0,p_-)$.
The second inverse participation ratio is thus
\begin{align}
 I_2^{\mathrm{comp}} &:=\sum_{x=1}^{\mathcal{D}}q_x^2=p_+^2+p_-^2 \notag\\
 &=\frac14\left[(1+u)^2+(1-u)^2\right]
 =\frac12(1+u^2),
 \qquad u=\operatorname{sech}(2\gamma) \notag\\
 &=1-\frac12\tanh^2(2\gamma).
 \label{eq:IPRexplicit}
\end{align}
On the other hand, Eq.~\eqref{eq:rhoAexplicit} gives
$\Tr[(\rho_A^R)^2]=p_+^2+p_-^2$.
Therefore the computational-basis IPR and the second R\'enyi entropy obey
\begin{equation}
 I_2^{\mathrm{comp}}
 =\Tr[(\rho_A^R)^2]
 =\exp[-S_A^{(2,R)}].
 \label{eq:IPRpurityrelation}
\end{equation}
This equality is not basis independent.  It holds here because the only
occupied computational product states are precisely the two Schmidt product
states.  Under a generic global basis rotation the IPR changes, whereas the
Schmidt probabilities and the entanglement entropy do not.
To see this, consider the global rotation into the Bell basis $\{\vert{}\Phi^\pm\rangle, \vert{}\Psi^\pm\rangle\}$. In this basis, the state reads $\vert{}\tilde{R}_{00}\rangle = \frac{a_\gamma+b_\gamma}{\sqrt{2}}\vert{}\Phi^+\rangle + \frac{a_\gamma-b_\gamma}{\sqrt{2}}\vert{}\Phi^-\rangle$. The corresponding inverse participation ratio becomes $I_2^{\mathrm{Bell}} = \left(\frac{(a_\gamma+b_\gamma)^2}{2}\right)^2 + \left(\frac{(a_\gamma-b_\gamma)^2}{2}\right)^2 = \frac{1+C_\gamma^2}{2} = 1 - \frac{u^2}{2}$.
As entanglement increases ($\vert{}\gamma\vert{} \to \infty, u \to 0$), the computational IPR decreases from 1 to 0.5, while the Bell-basis IPR increases from 0.5 to 1. This shows that the IPR is a basis-dependent delocalization metric, and its algebraic coincidence with reduced purity is restricted exclusively to the Schmidt basis.

In the limit $\abs{\gamma}\to\infty$, the squared amplitudes satisfy
$p_+ \to 0.5, \quad p_- \to 0.5$,
so that $I_2^{\mathrm{comp}}\longrightarrow\frac12$, and $S_A^{(2,R)}\longrightarrow\ln2$.
In this limit, $\vert{}\widetilde{R}_{00}\rangle \xrightarrow{\vert{}\gamma\vert{}\to\infty} \frac{1}{\sqrt{2}}\vert{}00\rangle + \frac{1}{\sqrt{2}}\vert{}11\rangle$,
and the probabilities $q_x = \vert{}\langle x \vert{} \widetilde{R}_{00} \rangle\vert{}^2= \text{Tr}(\vert{}x\rangle\langle x\vert{} \widetilde{R}_{00} \rangle\langle \widetilde{R}_{00} \vert{})$ are
$q_{00} = \left\vert{}\frac{1}{\sqrt{2}}\right\vert{}^2 = \frac{1}{2}$,
$q_{01} = 0$,
$q_{10} = 0$,
$q_{11} = \left\vert{}\frac{1}{\sqrt{2}}\right\vert{}^2 = \frac{1}{2}$.
Such that $I_2 = q_{00}^2 + q_{01}^2 + q_{10}^2 + q_{11}^2 = (0.5)^2 + 0 + 0 + (0.5)^2 = 0.5$ and the effective participation number is $D_{\text{eff}} = \frac{1}{I_2} = \frac{1}{0.5} = 2$.
This defines the maximally delocalized state accessible within the model.
The Haar-averaged IPR $\mathbb{E}_{\text{Haar}}[I_2] = 0.4$ serves as an exact baseline for maximal ergodic delocalization across the entire state space.
In undeformed limit $\gamma=0$ the state is localized on a single computational basis state with $q = (1, 0, 0, 0)$, yielding $I_2 = 1$ and an effective dimension $D_{\text{eff}} = 1/I_2 = 1$. This corresponds to the most localized, unentangled product state within the model.
At $\vert{}\gamma\vert{}\to\infty$, the canonical state has $I_2=0.5$,
above the complex-Haar average $0.4$ and the uniform-distribution lower
bound $0.25$.  This offset follows from its two-component support in the
computational basis.

The state is maximally entangled across $A|B$, but it still occupies only two
of the four global computational-basis vectors.  For comparison, a complex
Haar-random vector in dimension $\cD$ satisfies
\begin{equation}
\begin{aligned}
& \mathbb{E}_{\text{Haar}}[q_x^2] = \text{Tr}\Big[ (\vert{}x\rangle\langle x\vert{})^{\otimes 2} \mathbb{E}_{\mathrm{Haar}}\left[\Psi^{\otimes 2}\right] \Big] 
=  \frac{\text{Tr}\Big[ (\vert{}x\rangle\langle x\vert{})^{\otimes 2} I \Big] + \text{Tr}\Big[ (\vert{}x\rangle\langle x\vert{})^{\otimes 2} F_{AB} \Big]}{\mathcal{D}(\mathcal{D}+1)}= \frac{2}{\mathcal{D}(\mathcal{D}+1)},\\
& \mathbb{E}_{\text{Haar}} I_2^{\text{comp}} = \mathbb{E}_{\text{Haar}} \left[ \sum_{x=1}^{\mathcal{D}} q_x^2 \right] = \sum_{x=1}^{\mathcal{D}} \mathbb{E}_{\text{Haar}}[q_x^2]
 = \frac{2\mathcal{D}}{\mathcal{D}(\mathcal{D}+1)} = \frac{2}{\mathcal{D}+1},
 \label{eq:HaarIPR}
 \end{aligned}
\end{equation}
 where $x \in \{00, \, 01, \, 10, \, 11\}$, $q_x^2 = \Big[ \text{Tr}(\vert{}x\rangle\langle x\vert{} \Psi) \Big]^2 = \text{Tr}\Big[ (\vert{}x\rangle\langle x\vert{})^{\otimes 2} \Psi^{\otimes 2} \Big]$ and the 2nd moment/Weingarten integration reads $\mathbb{E}_{\mathrm{Haar}}\left[\Psi^{\otimes 2}\right] = \frac{I + F_{AB}}{\mathcal{D}(\mathcal{D}+1)}$.
Here $\text{Tr}\Big[(\vert{}x\rangle\langle x\vert{})^{\otimes 2} I\Big] = \langle x, x \vert{} I \vert{} x, x \rangle = \langle x\vert{}x\rangle \cdot \langle x\vert{}x\rangle = 1$ and $\text{Tr}\Big[ (\vert{}x\rangle\langle x\vert{})^{\otimes 2} F_{AB} \Big]=\langle x, x \vert{} F_{AB} \vert{} x, x \rangle = \langle x, x \vert{} x, x \rangle = 1$.
These results are independent of the basis state $\vert{}x\rangle$ because of
the spherical symmetry of Haar sampling.
At $\cD=4$, the latter value is $2/5$, below the limiting value $1/2$ in $I_2^{\mathrm{comp}}$. Thus maximal bipartite entanglement does not imply
full Hilbert-space anticoncentration or Porter--Thomas output statistics.  A
Haar-random realization is not exactly uniform over the four basis states;
Eq.~\eqref{eq:HaarIPR} is the ensemble average of its fluctuating output
probabilities \cite{Magni2025Anticoncentration}.

\subsection{Geometry of the degenerate right eigenspace}

The parameter dependence of an individual eigenvector inside a degenerate
eigenspace is basis dependent.  A basis-independent geometric diagnostic is
instead obtained from the orthogonal projector $Q_0^R$ onto the complete right
eigenspace.
Introduce $a=\cosh\gamma/\sqrt{\cosh(2\gamma)}$ and
$b=\sinh\gamma/\sqrt{\cosh(2\gamma)}$, with $a^2+b^2=1$, and the
orthonormal states
\begin{align}
 \ket{\psi_0}&=a\ket{00}+b\ket{11},
 &\ket{\chi_0}&=-b\ket{00}+a\ket{11},\notag\\
 \ket{\psi_1}&=a\ket{01}+b\ket{10},
 &\ket{\chi_1}&=-b\ket{01}+a\ket{10}.
 \label{eq:psichi}
\end{align}
\begingroup
This distinction is already visible within the present doublet.  A normalized
state in the degenerate right eigenspace can be written as
\begin{equation}
 |\Psi\rangle=\alpha|\psi_0\rangle+\beta|\psi_1\rangle,
 \qquad |\alpha|^2+|\beta|^2=1,
 \label{eq:GeneralDegenerateRightState}
\end{equation}
whose coefficient matrix in the computational product basis is
\begin{equation}
 C_\Psi=
 \begin{pmatrix}
  \alpha a&\beta a\\
  \beta b&\alpha b
 \end{pmatrix},
 \qquad
 \det C_\Psi=ab(\alpha^2-\beta^2).
 \label{eq:DegenerateSubspaceCoefficientMatrix}
\end{equation}
Thus the same degenerate eigenspace contains separable eigenvectors whenever
$\alpha=\pm\beta$.  For example,
\begin{equation}
 |\psi_0\rangle+|\psi_1\rangle
 =(a|0\rangle+b|1\rangle)_A\otimes(|0\rangle+|1\rangle)_B.
 \label{eq:SeparableVectorInDegenerateSubspace}
\end{equation}
The product-to-Bell interpolation therefore characterizes the transported
canonical eigenbasis, not every vector in the degenerate eigenspace.
\endgroup
Here $\ket{\psi_0}$ and $\ket{\psi_1}$ span $\mathcal E_0^R$, while each
$\ket{\chi_\mu}$ is orthogonal to the entire degenerate right eigenspace.  In
the ordered computational basis, the orthogonal projector is explicitly
\begin{equation}
 Q_0^R=\sum_{\mu=0}^{1}\ket{\psi_\mu}\bra{\psi_\mu}
 =\begin{pmatrix}
 a^2&0&0&ab\\
 0&a^2&ab&0\\
 0&ab&b^2&0\\
 ab&0&0&b^2
 \end{pmatrix}.
 \label{eq:Qexplicit}
\end{equation}

To differentiate the normalized states without carrying derivatives of their
normalization factors separately, write
\begin{equation}
 a=\cos\theta(\gamma),\qquad b=\sin\theta(\gamma),
 \qquad \tan\theta(\gamma)=\tanh\gamma.
 \label{eq:theta}
\end{equation}
Differentiating the last relation yields
\begin{align}
 \sec^2\theta\,\partial_\gamma\theta
 &=\operatorname{sech}^2\gamma,\notag\\
 \partial_\gamma\theta
 &=\frac{\operatorname{sech}^2\gamma}
 {1+\tanh^2\gamma}
 =\frac{1}{\cosh(2\gamma)}
 =\operatorname{sech}(2\gamma).
 \label{eq:thetaderivative}
\end{align}
Equations~\eqref{eq:psichi} and \eqref{eq:thetaderivative} then give
\begin{equation}
 \partial_\gamma\ket{\psi_\mu}
 =\operatorname{sech}(2\gamma)\ket{\chi_\mu},
 \qquad \mu=0,1.
 \label{eq:psiderivative}
\end{equation}
Since $\braket{\psi_\mu}{\chi_\nu}=0$, the pure-state Fubini--Study
susceptibility of either natural right eigenvector is
\begin{align}
 \chi_\gamma^R
 &:=\braket{\partial_\gamma\psi_\mu}
 {\partial_\gamma\psi_\mu}
 -\abs{\braket{\psi_\mu}{\partial_\gamma\psi_\mu}}^2\notag\\
 &=\operatorname{sech}^2(2\gamma).
 \label{eq:rightFS}
\end{align}

The same result follows directly at the subspace level.  Differentiating
Eq.~\eqref{eq:Qexplicit} in its spectral form gives
\begin{equation}
 \partial_\gamma Q_0^R
 =\theta'\sum_{\mu=0}^{1}
 \left(\ket{\chi_\mu}\bra{\psi_\mu}
 +\ket{\psi_\mu}\bra{\chi_\mu}\right),
 \qquad \theta'=\operatorname{sech}(2\gamma).
 \label{eq:Qderivative}
\end{equation}
Orthogonality between the two $\mu$ sectors removes all cross terms.  For each
$\mu$,
\begin{align}
 &\left(\ket{\chi_\mu}\bra{\psi_\mu}
 +\ket{\psi_\mu}\bra{\chi_\mu}\right)^2\notag\\
 &\hspace{2cm}=\ket{\chi_\mu}\bra{\chi_\mu}
 +\ket{\psi_\mu}\bra{\psi_\mu},
\end{align}
whose trace is $2$.  For the present rank-$r=2$ subspace, in which both basis
vectors have the same $\theta$ dependence, this establishes
\begin{equation}
 \Tr[(\partial_\gamma Q_0^R)^2]
 =2r(\theta')^2.
 \label{eq:Qtracesquare}
\end{equation}
With $r=2$, the normalized Grassmannian, or subspace Fubini--Study, metric is
therefore
\begin{equation}
 g_{\gamma\gamma}^{\mathrm{sub}}
 :=\frac{1}{2r}\Tr[(\partial_\gamma Q_0^R)^2]
 =\operatorname{sech}^2(2\gamma).
 \label{eq:subspacemetric}
\end{equation}

This result can also be checked from the generator
$G=\sigma_x\otimes\sigma_x$.  For either normalized right state,
\begin{equation}
\begin{aligned}
& \langle G\rangle:= \langle \widetilde{R}_{00} \vert{} G \vert{} \widetilde{R}_{00} \rangle
 = \frac{1}{\cosh(2\gamma)} \left( \cosh\gamma \langle 00\vert{} + \sinh\gamma \langle 11\vert{} \right) \left( \cosh\gamma \vert{}11\rangle + \sinh\gamma \vert{}00\rangle \right)\\
& =2ab=\frac{2\cosh\gamma\sinh\gamma}{\cosh(2\gamma)}
 =\tanh(2\gamma),
 \qquad \langle G^2\rangle=1,
 \end{aligned}
\end{equation}
and hence
$\operatorname{Var}(G)=\langle G^2\rangle-\langle G\rangle^2
=1-\tanh^2(2\gamma)=\operatorname{sech}^2(2\gamma)$.
The geometric susceptibility is largest at $\gamma=0$ and decays to zero as
$\abs{\gamma}\to\infty$, whereas the condition number in
Eq.~\eqref{eq:kappa} grows exponentially.

It is important not to interpret Eq.~\eqref{eq:subspacemetric} as evidence of
quantum chaos.  The present deformation is exactly isospectral:
\begin{equation}
 \partial_\gamma H_\gamma=[G,H_\gamma],
 \qquad
 \bra{L_m}\partial_\gamma H_\gamma\ket{R_n}
 =(\lambda_n-\lambda_m)\bra{L_m}G\ket{R_n}.
 \label{eq:isospectralderivative}
\end{equation}
Thus the energy difference appearing in an adiabatic-gauge-potential matrix
element cancels rather than producing a small-gap enhancement.  Isospectral
symmetry-generated directions are correspondingly excluded when the
adiabatic gauge potential is used as a chaos diagnostic
\cite{Pandey2020}.  Here Eqs.~\eqref{eq:rightFS} and
\eqref{eq:subspacemetric} quantify only the geometry of the non-unitarily
deformed right eigenspace.

\subsection{PPT negativity as an operational two-qubit witness}

For a positive two-qubit right-state density matrix $\rho^R$, partial transposition of
subsystem $B$ is defined in a product basis by
\begin{equation}
 \left(\rho^R\right)^{T_B}
 =\sum_{i,j,k,l}\rho_{ij,kl}^R
 \ket{i}\bra{k}\otimes\ket{l}\bra{j}.
 \label{eq:ppt}
\end{equation}
The PPT criterion is necessary and sufficient for separability in a
$2\times2$ system \cite{Peres1996}.  For either normalized state in
Eq.~\eqref{eq:Rdeg}, the spectrum of the partial transpose is
\begin{equation}
 \operatorname{spec}\!\left[(\rho^R)^{T_B}\right]
 =\left\{p_+,p_-,\sqrt{p_+p_-},-\sqrt{p_+p_-}\right\}.
 \label{eq:pptspectrum}
\end{equation}
It follows that the minimum eigenvalue and negativity are
\begin{equation}
 \lambda_{\min}^{T_B}(\gamma)
 =-\frac12\abs{\tanh(2\gamma)},
 \qquad
 \mathcal{N}_{\rm PPT}(\gamma)
 =\frac12\abs{\tanh(2\gamma)}.
 \label{eq:negativity}
\end{equation}
Thus the PPT witness grows monotonically from zero to its maximally entangled
value $1/2$, consistently with Eqs.~\eqref{eq:exactSvN} and
\eqref{eq:exactS2}.  Unlike the pure-state entropy, the same construction can
be applied to a noisy or incoherently mixed right-state density matrix.  Such a
two-qubit reduction can be reconstructed from Pauli correlators as
\begin{equation}
 \rho_{AB}^R=\frac14\sum_{\alpha,\beta=0}^{3}
 \left\langle\sigma_A^\alpha\otimes\sigma_B^\beta\right\rangle
 \sigma_A^\alpha\otimes\sigma_B^\beta,
 \label{eq:paulitomography}
\end{equation}
after which Eq.~\eqref{eq:ppt} gives a direct entanglement test
\cite{Baul2026}.  This test applies to the positive right-state density, not
to the generally non-positive biorthogonal reduction
Eq.~\eqref{eq:biodensity}.

\subsection{Non-local magic from the Schmidt spectrum}

For a pure qubit bipartition, local unitaries map the state to its
Schmidt-gauged representative
\begin{equation}
 \ket{\psi}_{\rm Sch}=\sum_{x\in\mathbb F_2^m}
 \sqrt{p_x}\ket{x}_A\ket{x}_B,
 \qquad m=\min(m_A,m_B).
 \label{eq:schmidtgauge}
\end{equation}
The second-order Schmidt-gauged non-local magic can be expressed through the
Walsh--Hadamard autocorrelations of the ordered Schmidt spectrum
\cite{Torre2026},
\begin{align}
 \mathcal{M}_{2}^{\rm Sch}
 &=-\log_2\left[2^{-m}
 \sum_{s,k\in\mathbb F_2^m}A_s(k)^4\right],
 \label{eq:magicgeneral}\\
 A_s(k)&=\sum_{x\in\mathbb F_2^m}
 \sqrt{p_xp_{x\oplus s}}(-1)^{k\cdot x}.
 \label{eq:walsh}
\end{align}
For the present $1\times1$ qubit bipartition this expression is exact and
depends only on $p_-$,
\begin{equation}
 \mathcal{M}_{2}^{\rm Sch}
 =1-\log_2\left[1+(1-2p_-)^4
 +16p_-^2(1-p_-)^2\right].
 \label{eq:magicp}
\end{equation}
Writing $u=\operatorname{sech}(2\gamma)$ and using
$p_-=(1-u)/2$, Eq.~\eqref{eq:magicp} reduces to the closed result
\begin{equation}
 \mathcal{M}_{2}^{\rm Sch}(\gamma)
 =-\log_2\left(1-u^2+u^4\right),
 \qquad u=\operatorname{sech}(2\gamma).
 \label{eq:magicgamma}
\end{equation}
\begingroup
The present state has Schmidt rank two.  The global computational-basis
optimality theorem for Schmidt rank not exceeding six therefore applies, so
the Schmidt representative realizes the minimum over unrestricted local
unitaries \cite{Sierant2026SpectralGeometry}.  Consequently,
\begin{equation}
 \mathcal M_2^{\rm NL}(\psi_\gamma)
 =\mathcal M_2^{\rm Sch}(\psi_\gamma)
 =-\log_2(1-u^2+u^4).
 \label{eq:ExactUnrestrictedNonlocalSRE}
\end{equation}
This upgrades Eq.~\eqref{eq:magicgamma} from a particular Schmidt-gauged
representative to the exact unrestricted two-qubit non-local stabilizer
R\'enyi entropy.  Equivalently, the single-dimer spectral function
$g(p)=1-4p(1-p)+16p^2(1-p)^2$ becomes
$g(p)=1-u^2+u^4$ after
$p(1-p)=(1-u^2)/4$.

The endpoint limit that formerly appeared in the many-mode spectral
discussion has a direct single-pair meaning without introducing a continuum
spectral density.  Since $C_\gamma^2=1-u^2$, expansion about the product-state
endpoint $u\to1^-$ gives
\begin{equation}
 \lim_{u\to1^-}
 \frac{\mathcal M_2^{\rm NL}(\psi_\gamma)}{1-u^2}
 =\frac{1}{\ln2},
 \qquad
 \mathcal M_2^{\rm NL}(\psi_\gamma)
 =\frac{C_\gamma^2}{\ln2}+O(C_\gamma^4)
 =\frac{4\gamma^2}{\ln2}+O(\gamma^4).
 \label{eq:WeakDeformationMagicScaling}
\end{equation}
Thus the exact non-local magic vanishes quadratically in both the concurrence
and the non-unitary deformation near the Hermitian product-state endpoint.

An independent fermionic-Gaussian analysis gives the entanglement-spectrum
upper bound
\begin{equation}
 M_{\alpha,>}^{\rm NL}
 =\frac{1}{1-\alpha}\sum_{j=1}^{\ell}
 \log_2\!\left[
 \frac{1+\nu_j^{2\alpha}+(1-\nu_j^2)^\alpha}{2}
 \right]
 \label{eq:TirritoGaussianSpectrumBound}
\end{equation}
for integer $\alpha>1$ \cite{Collura2026NonlocalMagic}.  In the present
single-pair problem $\nu_1=u$, and at $\alpha=2$ this expression becomes
\begin{align}
 M_{2,>}^{\rm NL}
 &=-\log_2\!\left[
 \frac{1+u^4+(1-u^2)^2}{2}
 \right]\notag\\
 &=-\log_2(1-u^2+u^4)
 =\mathcal M_2^{\rm Sch}.
 \label{eq:TirritoSingleModeAgreement}
\end{align}
Equation~\eqref{eq:TirritoSingleModeAgreement} independently reproduces the
single-mode function in Eq.~\eqref{eq:magicgamma} and identifies it as the
one-mode limit of a general Gaussian entanglement-spectrum construction.  In
general subsystem size, its optimality is supported by a local-minimum
analysis and simulated annealing; the quantity is therefore retained as an
upper bound rather than identified with unrestricted non-local magic
\cite{Collura2026NonlocalMagic}.
\endgroup
This quantity vanishes both at $\gamma=0$, where the right eigenstate is a
product state, and at $\abs{\gamma}\to\infty$, where its Schmidt spectrum is
perfectly flat.  It reaches
\begin{equation}
 \mathcal{M}_{2,\max}^{\rm Sch}=\log_2\frac43
 \label{eq:magicmax}
\end{equation}
when $u^2=1/2$, or
$\abs{\gamma}=\tfrac12\operatorname{arcosh}\sqrt2$.  Hence the amount of
entanglement grows monotonically while non-local magic is nonmonotonic.  This
provides information that neither the eigenvalue degeneracy nor the entropy
alone contains.  It also obeys the general spectral bound
\begin{equation}
 \mathcal{M}_{2}^{\rm Sch}
 \le 2S_{2,\mathrm{bits}}^R,
 \qquad
 S_{2,\mathrm{bits}}^R=-\log_2\Tr[(\rho_A^R)^2]
 =\frac{S_A^{(2,R)}}{\ln2}.
 \label{eq:magicbound}
\end{equation}

A stronger simultaneous constraint has been established for an arbitrary
pure-state qubit bipartition in terms of the occupied Schmidt rank
$r_{\rm Sch}$: $[S_{2,\mathrm{bits}}]^2+[\mathcal M_2^{\rm NL}]^2
\le q^2$, with $q=\lceil\log_2 r_{\rm Sch}\rceil$
\cite{Giampaolo2026Budget}.  Here $r_{\rm Sch}=2$ away from the product
endpoint, hence $q=1$.  Within the present family one can check this bound
without further optimization.  Indeed,
\begin{equation}
 \left(1+u^2\right)\left(1-u^2+u^4\right)=1+u^6\ge1,
 \qquad
 S_{2,\mathrm{bits}}^R=-\log_2\frac{1+u^2}{2},
 \label{eq:BudgetSinglePairIdentity}
\end{equation}
which implies the stronger rank-two inequality
\begin{equation}
 S_{2,\mathrm{bits}}^R+\mathcal M_2^{\rm NL}\le1,
 \qquad
 \left(S_{2,\mathrm{bits}}^R\right)^2+
 \left(\mathcal M_2^{\rm NL}\right)^2\le1.
 \label{eq:BudgetSinglePairCheck}
\end{equation}
At $\gamma=0$ the occupied Schmidt rank becomes one ($q=0$), and both
resources vanish.  Equation~\eqref{eq:BudgetSinglePairCheck} is the
rank-two specialization of the universal resource budget above.

\subsection{Exact non-local magic range of the degenerate right eigenspace}
\label{sec:DegenerateSubspaceMagicRange}

The Schmidt-gauged calculation above concerns the canonical transported
state $|\psi_0\rangle$, but a multiplicity-two eigenvalue allows any
normalized superposition in Eq.~\eqref{eq:GeneralDegenerateRightState}.
The determinant of its coefficient matrix in
Eq.~\eqref{eq:DegenerateSubspaceCoefficientMatrix} immediately gives the
pure-state concurrence
\begin{equation}
 \mathcal C(\Psi)=2|\det C_\Psi|
 =T_\gamma|\alpha^2-\beta^2|,
 \qquad T_\gamma=|\tanh(2\gamma)|.
 \label{eq:SubspaceConcurrence}
\end{equation}
For an arbitrary normalized two-qubit pure state, the Schmidt probabilities
are $p_\pm=(1\pm\sqrt{1-\mathcal C^2})/2$.  The exact Schmidt-gauge
optimality in the present $1|1$ partition, applied to the spectrum of
$|\Psi\rangle$, therefore gives the \emph{unrestricted} non-local
stabilizer R\'enyi-2 entropy
\begin{equation}
 \mathcal M_2^{\rm NL}(\Psi)
 =-\log_2[1-\mathcal C(\Psi)^2+\mathcal C(\Psi)^4].
 \label{eq:SubspaceExactMagic}
\end{equation}
This equality relies on the established local-unitary optimality theorem
\cite{Sierant2026SpectralGeometry}; the new specialization here is its
application to \emph{every} right eigenvector in the same non-defective
eigenspace, rather than to one designated Schmidt pair.

As $|\alpha|^2+|\beta|^2=1$, the factor
$|\alpha^2-\beta^2|$ takes every value in $[0,1]$.  Thus
$0\le\mathcal C(\Psi)\le T_\gamma$, and the minimum of
Eq.~\eqref{eq:SubspaceExactMagic} on the right eigenspace is always zero.
It is attained by the product states with
$\alpha=\pm\beta=1/\sqrt2$.  The function
$-\log_2(1-C^2+C^4)$ reaches its absolute maximum
$\log_2(4/3)$ at $C^2=1/2$.  The intrinsic
\emph{maximal resource capacity} of this degenerate right eigenspace is
\begin{equation}
 \mathcal M_{2,\max}^{\mathcal E_0^R}(\gamma)
 :=\max_{\substack{|\Psi\rangle\in\mathcal E_0^R\\
                   \langle\Psi|\Psi\rangle=1}}
 \mathcal M_2^{\rm NL}(\Psi)
 =\begin{cases}
 -\log_2(1-T_\gamma^2+T_\gamma^4),
   & |\gamma|\le\gamma_*,\\
 \log_2(4/3),&|\gamma|\ge\gamma_*,
 \end{cases}
 \label{eq:SubspaceMaximumMagic}
\end{equation}
where
$\gamma_*=(1/2)\operatorname{arcosh}\sqrt2\simeq0.440687$.
For $|\gamma|\le\gamma_*$, a canonical basis vector
($\alpha=1,\beta=0$) attains the maximum.  For
$|\gamma|\ge\gamma_*$, one possible maximizing superposition is
\begin{equation}
 |\Psi_{\max}(\gamma)\rangle=
 \cos\theta_\gamma|\psi_0\rangle+
 \sin\theta_\gamma|\psi_1\rangle,
 \quad
 \theta_\gamma=\frac12\arccos
 \left[\frac{1}{\sqrt2\,T_\gamma}\right],
 \label{eq:SubspaceMaximizer}
\end{equation}
which obeys $\mathcal C(\Psi_{\max})=1/\sqrt2$.
The canonical-state magic decreases to zero in the Bell limit, whereas
the eigenspace maximum remains at $\log_2(4/3)$ above the threshold and
the minimum stays zero.  Both extrema are independent of the basis chosen
within the right eigenspace for the fixed physical bipartition.  The
saturation reflects the geometry of this finite-dimensional doublet.

\begingroup

\subsection{A Triangle-Criterion witness for mixed-state magic}

The exact pure-state non-local stabilizer R\'enyi entropy above is defined for
pure states.  A complementary
question is whether non-stabilizerness can still be certified when the right
state is affected by experimental noise.  The triangle criterion provides a
sufficient mixed-state magic witness and detects every pure multi-qubit magic
state \cite{Liu2026Triangle}.  For the present two-qubit family, an explicit
stabilizer triangle can be constructed without searching over the Clifford
group.

We next define the non-negative amplitudes
\begin{equation}
 a_\gamma=\frac{\cosh\gamma}{\sqrt{\cosh(2\gamma)}},
 \qquad
 b_\gamma=\frac{\abs{\sinh\gamma}}{\sqrt{\cosh(2\gamma)}}.
 \label{eq:triangleab}
\end{equation}
Their normalization follows explicitly from
\begin{align}
 a_\gamma^2+b_\gamma^2
 &=\frac{\cosh^2\gamma+\sinh^2\gamma}{\cosh(2\gamma)}
 \notag\\
 &=\frac{\cosh(2\gamma)}{\cosh(2\gamma)}
 =1.
 \label{eq:triangleabnormalization}
\end{align}
When $\gamma<0$, we apply a local Pauli
$Z_A=\sigma_z\otimes\mathbb{I}_B$ to absorb the negative sign of $b_\gamma$:
indeed,
$Z_A(a_\gamma|00\rangle-b_\gamma|11\rangle)
=a_\gamma|00\rangle+b_\gamma|11\rangle$ because
$Z_A|00\rangle=|00\rangle$ and $Z_A|11\rangle=-|11\rangle$.
Since $Z_A$ is a Clifford operation, this sign change does not
alter the magic of the state.  
The sign-corrected Schmidt-gauged state is therefore
\begin{equation}
 \ket{\psi_\gamma}
 =a_\gamma\ket{00}+b_\gamma\ket{11}.
 \label{eq:trianglepsi}
\end{equation}
Applying the Clifford gate $\operatorname{CNOT}_{A\rightarrow B}$ gives
\begin{align}
 \operatorname{CNOT}_{A\rightarrow B}\ket{\psi_\gamma}
 &=a_\gamma\operatorname{CNOT}\ket{00}
   +b_\gamma\operatorname{CNOT}\ket{11}\notag\\
 &=a_\gamma\ket{00}+b_\gamma\ket{10}\notag\\
 &=\left(a_\gamma\ket{0}+b_\gamma\ket{1}\right)_A
   \otimes\ket{0}_B.
 \label{eq:trianglecnot}
\end{align}
Thus the two-qubit state is mapped by a \emph{global} Clifford operation
to a single-qubit superposition tensored with a stabilizer ancilla.
The CNOT acts across the $A|B$ bipartition: it preserves the ordinary
(global, fixed-basis) stabilizer R\'enyi entropy, but is not of the free
local-unitary form $U_A\otimes U_B$ used in the definition of
$\mathcal M_2^{\rm NL}$.  Indeed, the output is a product state and
has zero non-local magic even when the input has nonzero non-local magic.
We use this transformation as a global-Clifford reduction for studying
stabilizer-resource detection and concentration, not as a free LOCC or
local-unitary conversion of non-local magic.

Consider the three two-qubit stabilizer states
\begin{equation}
 \ket{\tau_1}=\ket{+}_A\ket{0}_B,
 \qquad
 \ket{\tau_2}=\ket{1}_A\ket{0}_B,
 \qquad
 \ket{\tau_3}=\ket{+i}_A\ket{0}_B,
 \label{eq:trianglestates}
\end{equation}
where $\ket{+}=(\ket0+\ket1)/\sqrt2$ and
$\ket{+i}=(\ket0+i\ket1)/\sqrt2$.
The common $\ket0_B$ factor reduces their pairwise fidelities to
single-qubit overlaps.  Directly,
\begin{align}
 \abs{\braket{\tau_1}{\tau_2}}^2
 &=\abs{\braket{+}{1}}^2
 =\abs{\frac{1}{\sqrt2}}^2=\frac12,\notag\\
 \abs{\braket{\tau_1}{\tau_3}}^2
 &=\abs{\braket{+}{+i}}^2
 =\abs{\frac{1+i}{2}}^2=\frac12,\notag\\
 \abs{\braket{\tau_2}{\tau_3}}^2
 &=\abs{\braket{1}{+i}}^2
 =\abs{\frac{i}{\sqrt2}}^2=\frac12.
 \label{eq:triangleoverlaps}
\end{align}
Hence the three states form a stabilizer triangle with
Their pairwise overlaps satisfy
$\abs{\braket{\tau_i}{\tau_j}}^2=1/2$ for $i\ne j$, as required by the
triangle criterion.

Let $\ket{\psi_\gamma'}=(a_\gamma\ket0+b_\gamma\ket1)_A\ket0_B$ and
$\rho_\gamma'=\ket{\psi_\gamma'}\bra{\psi_\gamma'}$ denote the
Clifford-transformed state.  Its fidelity with
$\ket{\tau_1}$ is
\begin{align}
 F_1
 &=\Tr(\rho_\gamma'\ket{\tau_1}\bra{\tau_1})
 =\abs{\braket{\tau_1}{\psi_\gamma'}}^2\notag\\
 &=\abs{\frac{a_\gamma+b_\gamma}{\sqrt2}}^2
 =\frac{(a_\gamma+b_\gamma)^2}{2}.
 \label{eq:triangleF1}
\end{align}
Similarly,
\begin{align}
 F_2
 &=\abs{\braket{\tau_2}{\psi_\gamma'}}^2
 =\abs{b_\gamma}^2
 =b_\gamma^2,
 \label{eq:triangleF2}\\
 F_3
 &=\abs{\braket{\tau_3}{\psi_\gamma'}}^2
 =\abs{\frac{a_\gamma-i b_\gamma}{\sqrt2}}^2\notag\\
 &=\frac{a_\gamma^2+b_\gamma^2}{2}
 =\frac12.
 \label{eq:triangleF3}
\end{align}
The amount by which this stabilizer triangle is violated is therefore
\begin{align}
 \Delta_\triangle(\gamma)
 &:=F_1-F_2-F_3\notag\\
 &=\frac{(a_\gamma+b_\gamma)^2}{2}
   -b_\gamma^2-\frac12\notag\\
 &=\frac{a_\gamma^2+2a_\gamma b_\gamma+b_\gamma^2
   -2b_\gamma^2-(a_\gamma^2+b_\gamma^2)}{2}\notag\\
 &=b_\gamma(a_\gamma-b_\gamma).
 \label{eq:trianglegap}
\end{align}
For every finite $0<\abs{\gamma}<\infty$, one has
$0<b_\gamma<a_\gamma$ because $\abs{\sinh\gamma}<\cosh\gamma$.
Consequently, $\Delta_\triangle(\gamma)>0$, or equivalently
$F_1>F_2+F_3$.
The triangle criterion then proves that the state is not a convex mixture of
stabilizer states.  At $\gamma=0$, one has $b_\gamma=0$ and
$\ket{\psi_\gamma}=\ket{00}$, which is a product stabilizer.  In the opposite
limit, $a_\gamma=b_\gamma=1/\sqrt2$ and the state approaches a Bell
stabilizer.  Thus $\Delta_\triangle$ vanishes at precisely the same two
endpoints as $\mathcal M_2^{\rm Sch}$ in
Eq.~\eqref{eq:magicgamma}.

The same calculation yields an explicit white-noise threshold.  Consider
\begin{equation}
 \rho_{\gamma,v}
 =v\ket{\psi_\gamma}\bra{\psi_\gamma}
 +(1-v)\frac{\id_4}{4},
 \qquad 0\le v\le1.
 \label{eq:whitenoisestate}
\end{equation}
Since the maximally mixed state is invariant under the preceding Clifford
transformation, the three noisy fidelities are
$F_i(v)=vF_i+(1-v)/4$ for $i=1,2,3$.
The noisy Triangle inequality is
\begin{align}
 F_1(v)&>F_2(v)+F_3(v),\notag\\
 vF_1+\frac{1-v}{4}
 &>v(F_2+F_3)+\frac{1-v}{2},\notag\\
 v(F_1-F_2-F_3)&>\frac{1-v}{4},\notag\\
 v\Delta_\triangle(\gamma)&>\frac{1-v}{4}.
 \label{eq:noisytriangleintermediate}
\end{align}
Multiplying by four and collecting the terms proportional to $v$ gives
\begin{align}
 4v\Delta_\triangle&>1-v,\notag\\
 v\left[1+4\Delta_\triangle\right]&>1.
\end{align}
Therefore this particular stabilizer triangle certifies mixed-state magic
whenever
\begin{equation}
 v>\frac{1}{1+4b_\gamma(a_\gamma-b_\gamma)}.
 \label{eq:trianglenoisethreshold}
\end{equation}
This is a sufficient condition for the noisy two-qubit state.  The Triangle
Criterion detects all single-qubit mixed magic and all multi-qubit pure magic,
but it is not a necessary test for arbitrary multi-qubit mixed states
\cite{Liu2026Triangle}.

The point of greatest noise tolerance can also be obtained analytically.
Write $a_\gamma=\cos\theta$ and $b_\gamma=\sin\theta$, with
$0\le\theta\le\pi/4$.
Then
\begin{align}
 \Delta_\triangle(\theta)
 &=\sin\theta(\cos\theta-\sin\theta)\notag\\
 &=\frac12\sin(2\theta)-\frac12[1-\cos(2\theta)]\notag\\
 &=\frac12\left[\sin(2\theta)+\cos(2\theta)-1\right].
 \label{eq:triangletheta}
\end{align}
Differentiation gives
$d\Delta_\triangle/d\theta=\cos(2\theta)-\sin(2\theta)$.  The stationary
point in the allowed interval satisfies $\tan(2\theta)=1$, hence
$\theta=\pi/8$.  Since
$1-2b_\gamma^2=a_\gamma^2-b_\gamma^2=\cos(2\theta)=u$,
this point corresponds to $u=1/\sqrt2$, exactly the maximum of
$\mathcal M_2^{\rm Sch}$ in Eq.~\eqref{eq:magicmax}.  At this point,
$\Delta_{\triangle,\max}=(\sqrt2-1)/2$, and
Eq.~\eqref{eq:trianglenoisethreshold} becomes
$v_{\min}=1/(2\sqrt2-1)\simeq0.547$.

The triangle observable can also serve as a seed in the recent
witness-expansion framework, which averages a linear witness over a
free-unitary orbit to construct nonlinear, multi-copy resource tests
\cite{Tang2026WitnessExpansion}.  In particular, the fourth-order expansion
connects triangle witnesses to stabilizer-R\'enyi diagnostics and can improve
mixed-state detection beyond a single fixed triangle.  This supplies a
systematic future extension of Eqs.~\eqref{eq:trianglegap}--
\eqref{eq:trianglenoisethreshold}, but it does not change their analytic
sufficient threshold and generally requires collective observables on several
copies.  The same framework also treats fermionic non-Gaussianity, whose free
set is the convex hull of pure fermionic Gaussian states; that resource is
distinct from the non-local stabilizer magic carried by the pure Gaussian
state considered here.

\subsection{Two-test verification in the Schmidt and mutually unbiased bases}

The analytic Schmidt structure permits direct verification of the prepared
right eigenstate without complete two-qubit tomography.  The
Schmidt-decomposition protocol \cite{Li2026Verification,Li2} uses one test
in the Schmidt basis and a second test in a mutually unbiased basis (MUB).
For finite $\gamma\ne0$, use the normalized coefficients $a$ and $b$ in
Eq.~\eqref{eq:R00normalized}.  They satisfy $a^2+b^2=1$, including when $b$ is
negative.  Define the conditional states of subsystem $B$ by
$\ket{\phi_\pm}_B=a\ket{0}_B\pm b\ket{1}_B$.
Their normalization is
\begin{align}
 \braket{\phi_\pm}{\phi_\pm}
 &=a^2+b^2
 \pm ab\braket{0}{1}
 \pm ab\braket{1}{0}\notag\\
 &=a^2+b^2=1.
 \label{eq:conditionalnormalization}
\end{align}

Using $\ket0=(\ket++\ket-)/\sqrt2$ and
$\ket1=(\ket+-\ket-)/\sqrt2$, the target state can be rewritten step by step as
\begin{align}
 \ket{\widetilde R_{00}}
 &=a\ket{0}_A\ket{0}_B+b\ket{1}_A\ket{1}_B\notag\\
 &=\frac{a}{\sqrt2}(\ket+_A+\ket-_A)\ket0_B
 +\frac{b}{\sqrt2}(\ket+_A-\ket-_A)\ket1_B\notag\\
 &=\frac{1}{\sqrt2}\left[
 \ket+_A(a\ket0_B+b\ket1_B)
 +\ket-_A(a\ket0_B-b\ket1_B)\right]\notag\\
 &=\frac{1}{\sqrt2}\left(
 \ket{+}_A\ket{\phi_+}_B
 +\ket{-}_A\ket{\phi_-}_B\right).
 \label{eq:mubdecomposition}
\end{align}

The first accepting projector tests the Schmidt-basis correlations,
\begin{equation}
 P_0=\ket{00}\bra{00}+\ket{11}\bra{11}.
 \label{eq:verificationP0}
\end{equation}
Acting on the target gives
\begin{align}
 P_0\ket{\widetilde R_{00}}
 &=\ket{00}\braket{00}{\widetilde R_{00}}
 +\ket{11}\braket{11}{\widetilde R_{00}}\notag\\
 &=a\ket{00}+b\ket{11}
 =\ket{\widetilde R_{00}}.
 \label{eq:P0pass}
\end{align}
The second accepting projector tests the MUB decomposition,
\begin{equation}
 P_1
=\ket{+}_{A}\bra{+}\otimes\ket{\phi_+}_{B}\bra{\phi_+}
+\ket{-}_{A}\bra{-}\otimes\ket{\phi_-}_{B}\bra{\phi_-}.
 \label{eq:verificationP1}
\end{equation}
Using Eq.~\eqref{eq:mubdecomposition}, orthogonality
$\braket{+}{-}=0$, and
$\braket{\phi_\pm}{\phi_\pm}=1$, one obtains
\begin{equation}
\begin{aligned}
 &P_1\ket{\widetilde R_{00}}
 = \frac{1}{\sqrt{2}} P_1 \left( \vert{}+\rangle_A \vert{}\phi_+\rangle_B + \vert{}-\rangle_A \vert{}\phi_-\rangle_B \right) \\ &= \frac{1}{\sqrt{2}} \left( \vert{}+\rangle_A \langle + \vert{} + \rangle_A \vert{}\phi_+\rangle_B \langle \phi_+ \vert{} \phi_+ \rangle_B + \vert{}-\rangle_A \langle - \vert{} - \rangle_A \vert{}\phi_-\rangle_B \langle \phi_- \vert{} \phi_- \rangle_B \right) \\
  &=\frac{1}{\sqrt2}\left(
 \ket+_A\ket{\phi_+}_B
 +\ket-_A\ket{\phi_-}_B\right)\notag\\
 &=\ket{\widetilde R_{00}}.
 \label{eq:P1pass}
\end{aligned}
\end{equation}
Thus $P_0\ket{\widetilde R_{00}}=P_1\ket{\widetilde R_{00}}
=\ket{\widetilde R_{00}}$.
Operationally, the first test measures both qubits in the computational
Schmidt basis and accepts only the correlated outcomes $00$ and $11$.  In the
second test, subsystem $A$ is measured in the $X$ basis; conditioned on the
outcome $+$ or $-$, subsystem $B$ is projected onto $\ket{\phi_+}$ or
$\ket{\phi_-}$, respectively.

Choosing the two tests with equal probability gives
\begin{equation}
 \Omega_{\rm ver}=\frac12(P_0+P_1).
 \label{eq:verificationoperator}
\end{equation}
To obtain its spectral gap, subtract the common target projector,
$\overline P_j=P_j-\ket{\widetilde R_{00}}\bra{\widetilde R_{00}}$ for
$j=0,1$.
For the bipartite Schmidt--MUB construction,
$\overline P_0$ and $\overline P_1$ are mutually orthogonal projectors
\cite{Li2026Verification}, so
$\overline P_0\overline P_1=\overline P_1\overline P_0=0$
and
\begin{align}
 \Omega_{\rm ver}
 -\ket{\widetilde R_{00}}\bra{\widetilde R_{00}}
 &=\frac12(\overline P_0+\overline P_1),\notag\\
\end{align}
Since the two projectors $\langle \psi \vert{} P_0 \vert{} \psi \rangle \le 1$ and $\langle \psi \vert{} P_1 \vert{} \psi \rangle \le 1$, the largest eigenvalue of $\Omega_{\rm ver}$ is one, associated with the target
state $\vert{}\widetilde R_{00}\rangle$.  
thus the state $\vert{}\psi\rangle$ resides entirely within the range of $P$,  $P_0 \vert{}\psi\rangle = \vert{}\psi\rangle$ and $P_1 \vert{}\psi\rangle = \vert{}\psi\rangle$, demanding that $\vert{}\psi\rangle$ passes both tests with $100\%$ probability. In the composite test design for quantum state verification (QSV), the intersection of the $+1$-eigenspaces (ranges) of the constituent test operators is constrained to be one-dimensional, spanned exclusively by the target state $\vert{}\widetilde R_{00}\rangle$,
$\text{range}(P_0) \cap \text{range}(P_1) = \text{span}\big(\vert{}\widetilde R_{00}\rangle\big)$. The target state $\vert{}\widetilde R_{00}\rangle$ is uniquely identified as the sole eigenstate corresponding to the maximum eigenvalue of $1$, thereby ensuring the robustness of the verification against any non-target state.
Its second-largest eigenvalue and spectral gap are
$ \beta(\Omega_{\rm ver})= \left\|
 \Omega_{\rm ver}
 -\ket{\widetilde R_{00}}\bra{\widetilde R_{00}}
 \right\|=\frac12$,
and
 $\nu(\Omega_{\rm ver})
 =1-\beta(\Omega_{\rm ver})=\frac12.
$. $\overline{P}_0$ and $\overline{P}_1$ are both positive semi-definite and orthogonal projection operators ($\overline{P}_0^{\dag} = \overline{P}_0, \overline{P}_0^2 = \overline{P}_0, \overline{P}_1^{\dag} = \overline{P}_1, \overline{P}_1^2 = \overline{P}_1$) and since $(\overline{P}_0 + \overline{P}_1)^2 = \overline{P}_0^2 + \overline{P}_0 \overline{P}_1 + \overline{P}_1 \overline{P}_0 + \overline{P}_1^2 = \overline{P}_0 +  \overline{P}_1$, $(\overline{P}_0 + \overline{P}_1)^2$ is also a orthogonal projection operator, thus $\Vert{}\overline{P}_0 + \overline{P}_1\Vert{} = 1$.
Since $\overline{P}_0 \vert{}\widetilde{R}_{00}\rangle = (P_0 - \vert{}\widetilde{R}_{00}\rangle\langle\widetilde{R}_{00}\vert{})\vert{}\widetilde{R}_{00}\rangle = \vert{}\widetilde{R}_{00}\rangle - \vert{}\widetilde{R}_{00}\rangle = 0=\overline{P}_1 \vert{}\widetilde{R}_{00}\rangle$, $\Omega_{{\rm ver}} = \vert{}\widetilde{R}_{00}\rangle\langle\widetilde{R}_{00}\vert{} + \frac{1}{2}\overline{P}_0 + \frac{1}{2}\overline{P}_1$ acts on target space $\text{span}(\vert{}\widetilde{R}_{00}\rangle)$ of target state as $\Omega \vert{}\widetilde{R}_{00}\rangle =\vert{}\widetilde{R}_{00}\rangle + \frac{1}{2}\overline{P}_0 \vert{}\widetilde{R}_{00}\rangle + \frac{1}{2}\overline{P}_1 \vert{}\widetilde{R}_{00}\rangle = \vert{}\widetilde{R}_{00}\rangle$. While the second-largest eigenvalue corresponds to the space $\text{range}(\overline{P}_0) \oplus \text{range}(\overline{P}_1)$.

Suppose a prepared state $\rho$ has target fidelity
$\bra{\widetilde R_{00}}\rho\ket{\widetilde R_{00}}\le1-\epsilon$.
The maximum probability of passing one test is then
$P_{\rm pass}^{(1)}=\operatorname{Tr}(\Omega\rho)
\le1-\nu\epsilon=1-\epsilon/2$.
For $N$ independent tests, the probability of passing all tests is bounded by
$p_{\rm pass}^{(N)}\le(1-\nu\epsilon)^N$.
Requiring this probability to be no larger than a significance level
$\delta$ gives
\begin{align}
 (1-\nu\epsilon)^N&\le\delta,\notag\\
 N\ln(1-\nu\epsilon)&\le\ln\delta,\notag\\
 N&\ge
 \frac{\ln\delta}{\ln(1-\nu\epsilon)}.
 \label{eq:verificationNexact}
\end{align}
The smallest sufficient integer is
$N_{\rm req}=\left\lceil\ln\delta/\ln(1-\nu\epsilon)\right\rceil$.
Using $-\ln(1-x)\ge x$ for $0<x<1$, $N (-\nu \epsilon) \le \ln \delta $, $N \nu \epsilon \ge \ln(\delta^{-1}) $, $N \ge \frac{\ln(\delta^{-1})}{\nu \epsilon}$. Thus
$\nu=1/2$ yields the convenient upper bound
$N_{\rm req}\le\lceil2\epsilon^{-1}\ln\delta^{-1}\rceil$.
At the product endpoint $\gamma=0$, the target can instead be verified
directly by local computational-basis measurements.  The protocol above
certifies fidelity to a known target value of $\gamma$; unlike tomography, it
does not reconstruct an unknown state and does not by itself determine its
entropy, PPT negativity, or magic.

\endgroup

\section{Haar-random entropy benchmarks}

For a normalized Haar-random state
$\ket{\psi}\in\mathbb{C}^{\cN}\otimes\mathbb{C}^{\cM}$ with
$\cN\le\cM$, the exact mean von Neumann entropy is Page's result
\cite{Page1993},
\begin{equation}
 \EE_{\mathrm{Haar}}S_A^{(\mathrm{vN})}
 =H_{\cN\cM}-H_{\cM}-\frac{\cN-1}{2\cM}
 \simeq\ln\cN-\frac{\cN}{2\cM},
 \label{eq:pageexact}
\end{equation}
where $H_n$ is the $n$th harmonic number and the last expression assumes
$1\ll\cN\le\cM$.  The Haar-averaged purity is
\cite{Lubkin1978,Zyczkowski2001}
\begin{equation}
 \EE_{\mathrm{Haar}}\Tr\rho_A^2
 =\frac{\cN+\cM}{\cN\cM+1}.
 \label{eq:purity}
\end{equation}
It gives the annealed second R\'enyi reference
\begin{align}
 \widetilde S_{A,\mathrm{Haar}}^{(2)}
 &:=-\ln\EE_{\mathrm{Haar}}\Tr\rho_A^2
 =\ln\frac{\cN\cM+1}{\cN+\cM}
 \label{eq:S2annealedexact}\\
 &\simeq\ln\cN-
 \ln\left[1+\frac{1}{\cM}
 \left(\cN-\frac{1}{\cN}\right)\right].
 \label{eq:S2approx}
\end{align}
The approximation in the second line follows at large $\cM$ and is the
standard large-environment form \cite{Kim2024}.  Strictly,
$\EE[-\ln\Tr\rho_A^2]\ne-\ln\EE[\Tr\rho_A^2]$; the latter is a lower
bound by Jensen's inequality and becomes accurate when purity is concentrated.
Equations~\eqref{eq:pageexact}--\eqref{eq:S2approx} are standard Haar
benchmarks, so their established derivations are not repeated here.  They
apply to the positive right-state reduction, not to the biorthogonal operator
in Eq.~\eqref{eq:biodensity}.

\begingroup
In a many-body bosonic problem the benchmark must additionally respect global
constraints.  Mid-spectrum Bose--Hubbard eigenstates with conserved particle
number are compared with a Page ensemble restricted to a fixed $U(1)$ sector,
and their subleading $O(1)$ contribution depends on particle density and the
local bosonic cutoff; number-nonconserving models exhibit a different
subleading structure \cite{Medos2026BoseHubbard}.  Those fixed-particle-number
formulas cannot be substituted into the present two-qubit calculation because
$G=\sigma_x\otimes\sigma_x$ couples the zero- and two-excitation sectors and
explicitly breaks $U(1)$.  The transferable lesson is instead that Page and
purity references must be conditioned on every exact global symmetry before
they are used as eigenstate-typicality benchmarks.
\endgroup

\subsection{Complex versus real Haar sampling}

The benchmark in Eq.~\eqref{eq:purity} assumes a complex Haar-random state.
This assumption is consistent with Eq.~\eqref{eq:subspacesample}, where the
sampling vector is explicitly complex Gaussian.  If one instead restricts the
coefficients to be real in order to preserve the real structure of
$H_\gamma$, the relevant ensemble is the orthogonal, or real, Haar ensemble,
whose fourth moment and average purity are different
\cite{Magni2025RealClifford}.

To see the distinction directly, write a normalized bipartite state as
$\ket{\psi}=\sum_{a=1}^{\cN}\sum_{\alpha=1}^{\cM}
\psi_{a\alpha}\ket{a}_A\ket{\alpha}_B$, where $\cD=\cN\cM$.
For real coefficients, the reduced density matrix and its purity are
\begin{align}
 (\rho_A)_{ac}&=\sum_{\alpha=1}^{\cM}
 \psi_{a\alpha}\psi_{c\alpha},\notag\\
 \Tr\rho_A^2
 &=\sum_{a,c=1}^{\cN}\sum_{\alpha,\beta=1}^{\cM}
 \psi_{a\alpha}\psi_{c\alpha}
 \psi_{c\beta}\psi_{a\beta}.
 \label{eq:realpurityexpanded}
\end{align}
For a vector uniformly distributed on the real unit sphere $S^{\cD-1}$,
rotational invariance gives the fourth moment
\begin{equation}
 \EE_{\mathbb R}(\psi_i\psi_j\psi_k\psi_l)
 =\frac{\delta_{ij}\delta_{kl}
 +\delta_{ik}\delta_{jl}
 +\delta_{il}\delta_{jk}}
 {\cD(\cD+2)}.
 \label{eq:realfourthmoment}
\end{equation}
In Eq.~\eqref{eq:realpurityexpanded}, choose the composite indices
$i=(a,\alpha)$, $j=(c,\alpha)$, $k=(c,\beta)$, and $l=(a,\beta)$.
The three Kronecker contractions in Eq.~\eqref{eq:realfourthmoment} then give,
respectively,
\begin{align}
 \sum_{a,c,\alpha,\beta}\delta_{ij}\delta_{kl}
 &=\sum_{a,c,\alpha,\beta}\delta_{ac}
 =\cN\cM^2,\notag\\
 \sum_{a,c,\alpha,\beta}\delta_{ik}\delta_{jl}
 &=\sum_{a,c,\alpha,\beta}\delta_{ac}\delta_{\alpha\beta}
 =\cN\cM,\notag\\
 \sum_{a,c,\alpha,\beta}\delta_{il}\delta_{jk}
 &=\sum_{a,c,\alpha,\beta}\delta_{\alpha\beta}
 =\cN^2\cM.
 \label{eq:realcontractions}
\end{align}
Substitution into Eq.~\eqref{eq:realpurityexpanded} yields
\begin{align}
 \EE_{\mathbb R}\Tr\rho_A^2
 &=\frac{\cN\cM^2+\cN\cM+\cN^2\cM}
 {\cN\cM(\cN\cM+2)}\notag\\
 &=\frac{\cN+\cM+1}{\cN\cM+2}.
 \label{eq:realHaarpurity}
\end{align}

For comparison, in the complex ensemble the nonzero fourth moment has only
the two pairings
\begin{equation}
\mathbb{E}_{\mathrm{Haar}} \big[ \langle i, k \vert{} \Psi^{\otimes 2} \vert{} j, l \rangle \big]
=
 \EE_{\mathbb C}(\psi_i\psi_j^*\psi_k\psi_l^*)
 = \frac{\overbrace{\langle i, k \vert{} I \vert{} j, l \rangle}^{\delta_{ij}\delta_{kl}} + \overbrace{\langle i, k \vert{} F_{AB} \vert{} j, l \rangle}^{\delta_{il}\delta_{kj}}}{\mathcal{D}(\mathcal{D}+1)}=\frac{\delta_{ij}\delta_{kl}+\delta_{il}\delta_{kj}}
 {\cD(\cD+1)}.
 \label{eq:complexfourthmoment}
\end{equation}
where we note that $\langle i, k \vert{} \Psi^{\otimes 2} \vert{} j, l \rangle = \langle i \vert{} \Psi \vert{} j \rangle \cdot \langle k \vert{} \Psi \vert{} l \rangle 
= \psi_i \psi_j^* \psi_k \psi_l^*$, $\langle i, k \vert{} I \vert{} j, l \rangle = \langle i, k \vert{} (I_1 \otimes I_2) \vert{} j, l \rangle= \langle i \vert{} I_1 \vert{} j \rangle \cdot \langle k \vert{} I_2 \vert{} l \rangle = \delta_{ij} \delta_{kl}$,
$\langle i, k \vert{} F_{AB} \vert{} j, l \rangle = \big( \langle i\vert{}_1 \langle k\vert{}_2 \big) F_{AB} \big( \vert{}j\rangle_1 \vert{}l\rangle_2 \big)= \big( \langle i\vert{}_1 \langle k\vert{}_2 \big) \big( \vert{}l\rangle_1 \vert{}j\rangle_2 \big)= \langle i \vert{} l \rangle \cdot \langle k \vert{} j \rangle = \delta_{il} \delta_{kj}$.

The corresponding two contractions are $\cN\cM^2$ and $\cN^2\cM$, which
recover Eq.~\eqref{eq:purity} as
$\EE_{\mathbb C}\Tr\rho_A^2=(\cN+\cM)/(\cN\cM+1)$.
The extra Wick contraction in the real ensemble is the origin of the $+1$ in
the numerator and the replacement $+1\mapsto+2$ in the denominator of
Eq.~\eqref{eq:realHaarpurity}.  Thus a numerical implementation must state
whether its Gaussian samples are complex or real.  The complex sampling
specified in Eq.~\eqref{eq:subspacesample} requires no change to the Page and
purity benchmarks used in this work; Eq.~\eqref{eq:realHaarpurity} should be
used only if the sampling prescription is deliberately restricted to real
coefficients.  The real-Haar mean von Neumann entropy likewise differs from
Page's complex-Haar formula and can be obtained from the corresponding
$\beta=1$ induced eigenvalue distribution or evaluated numerically.

\section{Entropy deficits for a degenerate eigenspace}

An individual eigenvector is not invariant under a change of basis within a
degenerate eigenspace.  We therefore associate entanglement with an ensemble
of normalized states in the right eigenspace
$\mathcal{E}_0^R=\Span\{\ket{R_{0\mu}}:\mu=1,\ldots,r\}$.
Let $Q_0^R$ denote the orthogonal Euclidean projector onto this subspace.
Draw a complex Gaussian vector $\ket{z}\in\mathbb C^{\cD}$ and form
\begin{equation}
 \ket{\phi_0}
 =\frac{Q_0^R\ket{z}}{\norm{Q_0^R\ket{z}}}.
 \label{eq:subspacesample}
\end{equation}
Equation~\eqref{eq:subspacesample} defines the uniform measure on the unit
sphere of the right eigenspace and is unchanged by a unitary rotation of its
basis.

For samples $\ket{\phi_0}$, define
\begin{align}
 \overline S_0^{(\mathrm{vN})}
 &=\EE_{\phi_0}\left[-\Tr\rho_A(\phi_0)\ln\rho_A(\phi_0)\right],\\
 \widetilde S_0^{(2)}
 &:=-\ln\EE_{\phi_0}\left[\Tr\rho_A(\phi_0)^2\right].
\end{align}
The signed deficits from the full-space Haar references are
\begin{align}
 \Delta_{\mathrm{vN}}
 &=S_{\mathrm{Page}}(\cN,\cM)-\overline S_0^{(\mathrm{vN})},
 \label{eq:deficit1}\\
 \Delta_2^{\mathrm{ann}}
 &=\widetilde S_{A,\mathrm{Haar}}^{(2)}
 -\widetilde S_0^{(2)}.
 \label{eq:deficit2}
\end{align}
The order of averaging matters for the R\'enyi-2 quantity:
$-\ln\EE[P]$ is an annealed estimator, while
$\EE[-\ln P]$ is a different (quenched) average; Jensen's inequality
gives $\EE[-\ln P]\ge-\ln\EE[P]$.
We deliberately compare \emph{annealed} subspace and annealed full-space
purities, and use sample means of $P=\Tr\rho_A^2$ to estimate
$\widetilde S_0^{(2)}$.  The reported standard error is obtained by the
first-order delta method,
$\operatorname{SE}(\widetilde S_0^{(2)})\simeq
\operatorname{SD}(P)/(\sqrt{n_{\rm samp}}\,\overline P)$.
The deficits quantify the difference between subspace entropy statistics
and the corresponding full-space Haar benchmarks.  Large positive values
indicate reduced entanglement relative to those reference values and can
reflect restrictions imposed by the eigenspace and bipartition.  These
comparisons concern static entropy statistics.

The full-space Page values are universal functions of $\cN$ and $\cM$.
By contrast, the subspace averages depend on both the rank $r$ and the
orientation of $\mathcal E_0^R$ relative to the tensor-product structure.
There is therefore no universal replacement obtained merely by substituting
$r$ for $\cN\cM$ in Eqs.~\eqref{eq:pageexact} and \eqref{eq:purity}.  This is
precisely why Eq.~\eqref{eq:subspacesample} is needed for a rank-deficient
non-Hermitian sector.

For the two-qubit model, Eqs.~\eqref{eq:exactSvN} and \eqref{eq:exactS2} give
the entropy of either natural degenerate eigenvector.  The equal bipartition
Haar references are
\begin{equation}
 S_{\mathrm{Page}}(2,2)=\frac13,
 \qquad
 \widetilde S_{A,\mathrm{Haar}}^{(2)}(2,2)=\ln\frac54.
 \label{eq:twoqubithaar}
\end{equation}
These are ensemble averages rather than upper bounds: at large $\abs\gamma$,
the exact eigenvectors reach $\ln2$ and thus become more entangled than a
typical two-qubit Haar state.

\section{Numerical implementation in larger dimensions}

The larger-dimensional calculation is used to test two claims of the core
construction: similarity preserves the multiplicity of the selected
eigenvalue, whereas the orientation of its right eigenspace relative to the
fixed tensor-product decomposition can change its entanglement statistics.
For a chosen bipartition $\cN\times\cM$, the numerical protocol is:
\begin{enumerate}
 \item Construct $H_0$ with a prescribed multiplicity-$r$ eigenvalue and
 choose one fixed Hermitian nonlocal generator $G$ and set
 $S_\gamma=e^{\gamma G}$ and $H_\gamma=S_\gamma H_0S_\gamma^{-1}$.
 \item Verify at every sampled $\gamma$ that
 $\operatorname{nullity}(H_\gamma-\lambda_0 I)=r$.  This is a numerical check
 of Proposition~1, not a search for an exceptional point.
 \item Transport a basis $V_0$ of the reference eigenspace to
 $S_\gamma V_0$, orthonormalize its columns by QR decomposition, and denote
 the resulting matrix by $Q_\gamma$.
 \item Draw complex Gaussian vectors $z\in\mathbb C^r$, form
 $|\phi_0\rangle=Q_\gamma z/\|Q_\gamma z\|$, and compute the reduced state,
 von Neumann entropy, and sample purity for the annealed second R\'enyi quantity as in
 Eqs.~\eqref{eq:subspacesample}, \eqref{eq:SvNright}, and
 \eqref{eq:S2right}.
 \item Report the Monte Carlo von Neumann mean and standard error, the
 annealed R\'enyi-2 estimator $-\ln(\overline P)$ and its propagated
 standard error, together with Page/Haar-purity references, signed deficits,
 the condition number
 $\kappa_2(S_\gamma)$, and the numerical nullity.
\end{enumerate}

As a concrete high-dimensional illustration, we use one fixed
$\cN=\cM=4$ ($\cD=16$) realization with $r=4$, $\lambda_0=0$, and
$H_0=\operatorname{diag}(0,0,0,0,1,\ldots,12)$.  The reference subspace
is $|0\rangle_A\otimes\mathcal H_B$.  To specify the fixed generator
used in the numerical code, let $K_A$ and $K_B$ be independently sampled
$4\times4$ Hermitian Gaussian matrices and let $K_{AB}$ be an independent
$16\times16$ Hermitian Gaussian matrix.  Each is formed by Hermitian
symmetrization of a matrix of independent complex normal entries.  Define
\begin{equation}
 G_{\rm raw}=K_A\otimes K_B+\tfrac14K_{AB},\qquad
 G=\frac{G_{\rm raw}-\frac{\Tr G_{\rm raw}}{\cD}I_{\cD}}
 {\big\|G_{\rm raw}-\frac{\Tr G_{\rm raw}}{\cD}I_{\cD}\big\|_\infty}.
 \label{eq:numericalGeneratorDefinition}
\end{equation}
The randomly drawn matrices are held fixed throughout the parameter
sweep.  Averaging over 1500 complex Gaussian subspace samples at each of
25 values in $0\leq\gamma\leq1.8$ gives
$\overline S_0^{(\mathrm{vN})}=\widetilde S_0^{(2)}=0$ at $\gamma=0$ and hence
the maximal deficits $\Delta_{\mathrm{vN}}=S_{\mathrm{Page}}(4,4)=0.9224$
and $\Delta_2^{\mathrm{ann}}=\widetilde S_{A,\mathrm{Haar}}^{(2)}(4,4)=0.7538$.  At
$\gamma=1.8$, the fixed realization gives
$\overline S_0^{(\mathrm{vN})}=0.485\pm0.003$ and
$\Delta_{\mathrm{vN}}=0.437$.  The corresponding annealed R\'enyi-2
estimator computed from the sample mean purity is
$\widetilde S_0^{(2)}=0.313\pm0.003$, with
$\Delta_2^{\mathrm{ann}}=0.441$; the stated standard error is propagated
using the first-order delta method.
Over the same sweep the condition number reaches
$\kappa_2(S_\gamma)=26.58$, while the numerical nullity remains pinned at
$r=4$ with rank tolerance $10^{-8}$.  Because $G$ is Hermitian,
$\kappa_2(S_\gamma)=\exp[\gamma(g_{\max}-g_{\min})]$ independently monitors
conditioning; a nullity change would therefore indicate numerical precision
loss rather than a physical exceptional point.  The reported values
refer to this fixed realization.

\begin{figure}[H]
 \centering
\includegraphics[width=0.98\linewidth]{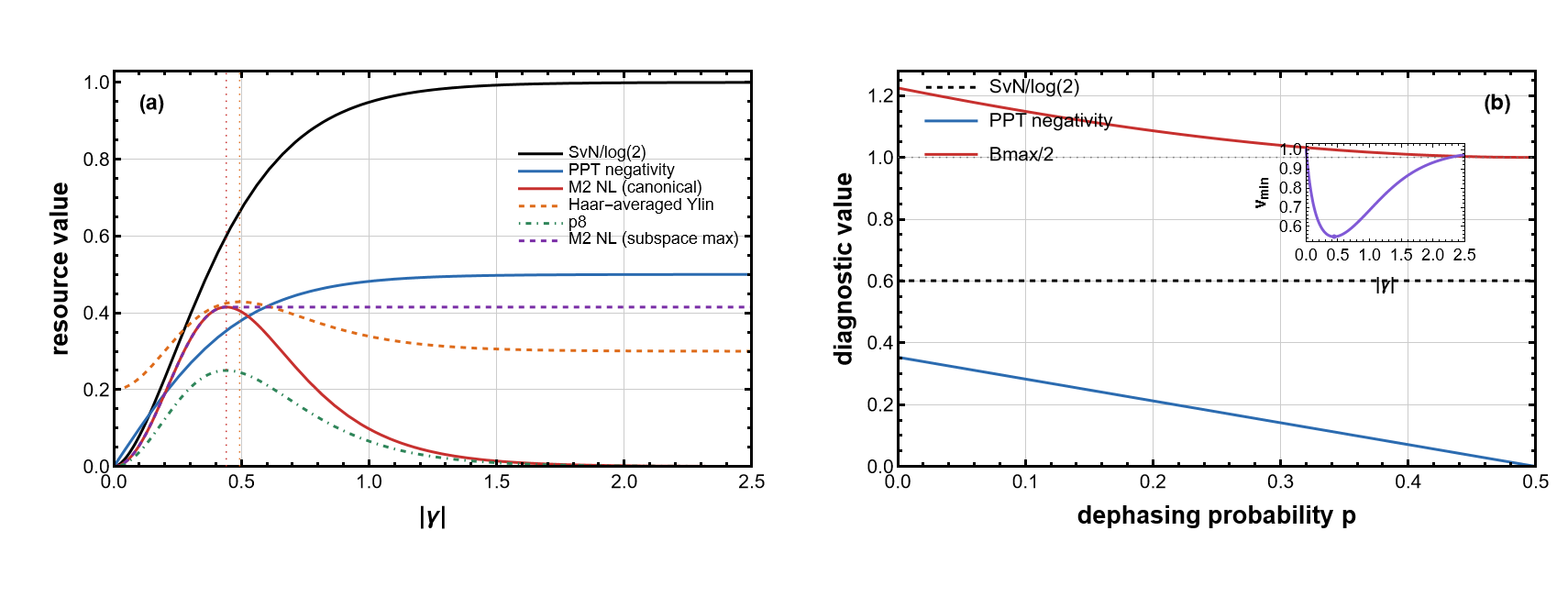}
\caption{Two-qubit resource competition, intrinsic resource capacity of the
non-defectively degenerate right eigenspace, and noise response.
(a) For the canonical right eigenstate $|\tilde{R}_{00}\rangle$,
pure-state entanglement entropy in bits $S_A^{(\mathrm{vN},R)}/\ln 2$ [Eq.~(\ref{eq:exactSvN})], PPT negativity $\mathcal{N}_{\mathrm{PPT}}$ [Eq.~(\ref{eq:negativity})], unrestricted non-local stabilizer R\'enyi entropy $\mathcal{M}_2^{\mathrm{NL}}$ [Eq.~(\ref{eq:ExactUnrestrictedNonlocalSRE})], locally Haar-averaged linear stabilizer entropy $\overline{Y^{\mathrm{lin}}}$ [Eq.~(\ref{eq:HaarquenchYlin})], and eight-copy CCZ achievable-protocol probability $p_8$ [Eq.~(\ref{eq:EightCopyCCZProbability})] as functions of $|\gamma|$. The dashed purple curve is the maximum
non-local magic over all normalized right eigenstates in the degenerate
eigenspace [Eq.~(\ref{eq:SubspaceMaximumMagic})]; it reaches the
$\log_2(4/3)$ plateau, unlike the decaying canonical-state curve.
Vertical dotted lines mark $u^2 = 1/2$ ($|\gamma_*| = \frac{1}{2}\operatorname{arcosh}\sqrt{2} \approx 0.441$) and $u^2 = 3/7$ ($|\gamma| = \frac{1}{2}\operatorname{arcosh}\sqrt{7/3} \approx 0.4933$), where $\mathcal{M}_2^{\mathrm{NL}}$ (along with $p_8$) and $\overline{Y^{\mathrm{lin}}}$ attain their respective maxima, demonstrating monotonic entanglement saturation versus nonmonotonic nonstabilizer behavior.
(b) Resource decay under the local dephasing channel $\mathcal{E}_p^{(B)}$ [Eq.~(\ref{eq:LocalChargeDephasingChannel})] evaluated at the maximal-magic state $u = C_\gamma = 1/\sqrt{2}$ ($|\gamma| = |\gamma_*|$): the reduced-state entropy $S_A^{(\mathrm{vN},R)}/\ln 2$ [Eq.~(\ref{eq:ReducedStateInvariantUnderDephasing})] remains invariant, whereas $\mathcal{N}_{\mathrm{PPT}}(p)$ [Eq.~(\ref{eq:DephasedNegativityConcurrence})] and the normalized CHSH value $\mathcal{B}_{\max}(p)/2$ [Eq.~(\ref{eq:DephasedCHSHMaximum})] decrease monotonically to the separable, Bell-local boundary $\mathcal{B}_{\max}/2 = 1$ (horizontal dotted line) at $p = 1/2$. The inset plots the sufficient white-noise visibility threshold $v_{\min}(\gamma) = [1 + 4b_\gamma(a_\gamma - b_\gamma)]^{-1}$ [Eq.~(\ref{eq:trianglenoisethreshold})] across the pure-state family, with the dot indicating its global minimum $v_{\min} = (2\sqrt{2}-1)^{-1} \approx 0.547$ at $|\gamma| = |\gamma_*|$.}
 \label{fig:ResourceCompetition}
\end{figure}

The high-dimensional calculation checks the subspace construction and
entropy benchmarks.  Figure~\ref{fig:ResourceCompetition} compares the
canonical non-local magic with the maximum over the degenerate right
eigenspace and displays their two-qubit response to noise.

\section{Operational extensions and resource structure}

\begingroup

\subsection{Pauli-spectrum, replica, and randomized-moment diagnostics}

\paragraph{Pauli moments and local Haar quench.}

The low-weight stabilizer R\'enyi entropy is a scalable diagnostic of many-body
localization and trainability, but it does not add an independent quantity in
the present two-qubit model \cite{Cao2025MBL}.  This can be seen
directly from the Pauli fourth moments.  With
$u=\operatorname{sech}(2\gamma)$ and $v_\gamma=\tanh(2\gamma)$,
the nonzero Pauli expectation values are
\begin{align}
 \langle II\rangle&=1,\qquad
 \langle ZZ\rangle=1,\notag\\
 \langle ZI\rangle
 &=a^2-b^2
 =\frac{\cosh^2\gamma-\sinh^2\gamma}{\cosh(2\gamma)}
 =u,\notag\\
 \langle IZ\rangle&=u,\notag\\
 \langle XX\rangle
 &=2ab
 =\frac{2\cosh\gamma\sinh\gamma}{\cosh(2\gamma)}
 =v_\gamma,\notag\\
 \langle YY\rangle&=-v_\gamma.
 \label{eq:paulimoments}
\end{align}
where $(Z \otimes Z)\vert{}00\rangle = \vert{}00\rangle$, $(Z \otimes Z)\vert{}11\rangle = \vert{}11\rangle$, $(Y \otimes Y)\vert{}00\rangle = -\vert{}11\rangle$, 
$(X \otimes X)\vert{}00\rangle = \vert{}11\rangle$, $(X \otimes X)\vert{}11\rangle = \vert{}00\rangle$, 
$(Y \otimes Y)\vert{}11\rangle = -\vert{}00\rangle$, $(X \otimes Y)\vert{}00\rangle = i\vert{}11\rangle, \quad (X \otimes Y)\vert{}11\rangle = -i\vert{}00\rangle$.
All other two-qubit Pauli expectations vanish.  Consequently,
the complete signed Pauli spectrum of the sign-corrected Schmidt state can be
recorded without introducing a new state parameter.  With
$C_\gamma=\sqrt{1-u^2}=|v_\gamma|$, define the empirical distributions over
the sixteen two-qubit Pauli strings by
\begin{align}
 \Pi_\gamma(x)
 &=\frac{10}{16}\delta(x)
 +\frac{2}{16}\delta(x-1)
 +\frac{2}{16}\delta(x-u)
 +\frac{1}{16}\delta(x-C_\gamma)
 +\frac{1}{16}\delta(x+C_\gamma),
 \label{eq:ExactSignedPauliSpectrum}\\
 \Pi_\gamma^{(2)}(y)
 &=\frac{10}{16}\delta(y)
 +\frac{2}{16}\delta(y-1)
 +\frac{2}{16}\delta(y-u^2)
 +\frac{2}{16}\delta(y-C_\gamma^2).
 \label{eq:ExactSquaredPauliSpectrum}
\end{align}
These finite spectra are the two-mode specialization of the Pauli-spectrum
description of fermionic Gaussian states \cite{Turkeshi2026PauliSpectrum}.
They also extend the fourth-moment calculation below to every real $q>1$.
Using the two-qubit dimension $D=4$, the stabilizer purity and its associated
fixed-basis stabilizer R\'enyi entropy are
\begin{align}
 \zeta_q(\gamma)
 &:=\frac{1}{4}\sum_{P\in\mathcal P_2}
 \left(\langle P\rangle^2\right)^q
 =\frac12\left[1+u^{2q}+(1-u^2)^q\right],
 \label{eq:ArbitraryOrderStabilizerPurity}\\
 M_q^{S}(\gamma)
 &=\frac{1}{1-q}\log_2\!\left\{
 \frac12\left[1+u^{2q}+(1-u^2)^q\right]
 \right\}.
 \label{eq:ArbitraryOrderStabilizerRenyi}
\end{align}
At $q=2$, Eq.~\eqref{eq:ArbitraryOrderStabilizerPurity} reduces to
$\zeta_2=1-u^2+u^4$, and Eq.~\eqref{eq:ArbitraryOrderStabilizerRenyi}
therefore reproduces Eq.~\eqref{eq:magicgamma}.  For every $q>1$, the maximum
occurs at $u^2=1/2$ and equals
\begin{equation}
 M_{q,\max}^{S}
 =\frac{1}{1-q}\log_2\!\left(\frac12+2^{-q}\right).
 \label{eq:ArbitraryOrderStabilizerMaximum}
\end{equation}
Only the order-two value is promoted below to the exact unrestricted
non-local stabilizer entropy; for $q\ne2$,
Eq.~\eqref{eq:ArbitraryOrderStabilizerRenyi} is retained as the SRE of this
fixed Schmidt representative.

The thermodynamic random-Gaussian conclusions---including the $q=2$
freezing transition, rare low-Majorana-weight dominance, exponential
high-order sampling cost, a finite traced-fraction threshold, and the
$\mathrm{SYK}_2$ ensemble correspondence---require Haar-random Gaussian
orbits and ensemble averaging and do not apply to this one-dimensional
two-mode manifold \cite{Turkeshi2026PauliSpectrum}.  Indeed,
$\rho_A=a_\gamma^2|0\rangle\langle0|+b_\gamma^2|1\rangle\langle1|$ is a
classical mixture of stabilizer states and has no local magic; this is fully
consistent with, but not representative of, the large random-multimode
ensemble.

Consequently,
\begin{align}
 \sum_{P\in\mathcal P_2}\langle P\rangle^4
 &=\langle II\rangle^4+\langle ZZ\rangle^4
 +\langle ZI\rangle^4+\langle IZ\rangle^4
 +\langle XX\rangle^4+\langle YY\rangle^4\notag\\
 &=2+2u^4+2v_\gamma^4.
\end{align}
Using $v_\gamma^2=1-u^2$ gives
\begin{align}
 2+2u^4+2v_\gamma^4
 &=2+2u^4+2(1-u^2)^2\notag\\
 &=2+2u^4+2(1-2u^2+u^4)\notag\\
 &=4(1-u^2+u^4).
 \label{eq:paulifourthsum}
\end{align}
With logarithms to base two, the $n=2$, $k=2$ full-weight specialization of
the normalized low-weight definition is
\begin{align}
 M_{2,2}^{\rm LW}
 &:=-\log_2\left[
 \frac{1}{16}\sum_{P\in\mathcal P_2}\langle P\rangle^4
 \right]\notag\\
 &=-\log_2\left[\frac{1-u^2+u^4}{4}\right]\notag\\
 &=2-\log_2(1-u^2+u^4)\notag\\
 &=2+\mathcal M_2^{\rm Sch}.
 \label{eq:lowweightsrerelation}
\end{align}
Thus the low-weight quantity differs from Eq.~\eqref{eq:magicgamma} only by an
additive normalization constant in this two-qubit setting.  Its localization
algorithm, disorder averaging, and barren-plateau analysis become relevant
only after extending the static construction to a genuine many-body
variational circuit.

The additive constant follows directly from the normalization of the
Pauli--Weyl phase-space probability distribution \cite{Crew2026Hybrid}.  For a
pure $n$-qubit state, define the
Pauli--Weyl probability distribution
\begin{equation}
 p_\rho(P)=2^{-n}\abs{\Tr(P\rho)}^2,
 \qquad P\in\mathcal P_n.
 \label{eq:pauliweylprobability}
\end{equation}
Its second R\'enyi entropy is
\begin{align}
 H_2(p_\rho)
 &=-\log_2\sum_{P\in\mathcal P_n}p_\rho(P)^2\notag\\
 &=-\log_2\left[
 2^{-2n}\sum_{P\in\mathcal P_n}\langle P\rangle^4
 \right]
  \\ &= -\log_2 (2^{-n}) - \log_2 \left[ 2^{-n} \sum_{P} \vert{}\text{Tr}(P\rho)\vert{}^4 \right] \\ &= n + M_2^S(\rho) .
 \label{eq:pauliweylH2}
\end{align}
The conventional second stabilizer R\'enyi entropy includes the shift by
$n$ bits,
\begin{align}
 M_2^{S}(\rho)
 &:=H_2(p_\rho)-n\notag\\
 &=-\log_2\left[
 2^{-n}\sum_{P\in\mathcal P_n}\langle P\rangle^4
 \right].
 \label{eq:standardSRE}
\end{align}
For $n=2$, substitution of Eq.~\eqref{eq:paulifourthsum} gives
\begin{align}
 M_2^{S}(\psi_\gamma)
 &=-\log_2\left[
 \frac14\,4(1-u^2+u^4)
 \right]\notag\\
 &=-\log_2(1-u^2+u^4)\notag\\
 &=\mathcal M_2^{\rm Sch}.
 \label{eq:standardSREequalsSchmidt}
\end{align}
For the deformed right eigenstate, the equality between the full stabilizer
R\'enyi entropy $M_2^S(\psi_\gamma)$ and the exact non-local value
$\mathcal M_2^{\rm NL}=\mathcal M_2^{\rm Sch}$ shows that the
non-stabilizerness of this doublet representative is entirely non-local.  The
single-qubit reductions contain no residual local magic; the resource is fixed
by the spectral organization of the bipartite Schmidt weights.
Consequently,
\begin{equation}
 M_{2,2}^{\rm LW}
 =H_2(p_{\psi_\gamma})
 =M_2^S(\psi_\gamma)+2
 =\mathcal M_2^{\rm Sch}+2.
 \label{eq:normalizationdictionary}
\end{equation}
Thus Eqs.~\eqref{eq:lowweightsrerelation} and
\eqref{eq:standardSREequalsSchmidt} are not competing definitions: the former
retains the unshifted Pauli-distribution entropy, whereas the standard
stabilizer R\'enyi entropy subtracts the two-bit reference constant.  

There is also a formal connection to magic injection.  Equation
\eqref{eq:trianglepsi} has the imperfect-Bell-pair form
$\ket{\psi_\gamma}=\cos\theta\ket{00}+\sin\theta\ket{11}$,
with
\begin{equation}
 \tan\theta
 =\frac{b_\gamma}{a_\gamma}
 =\frac{\abs{\sinh\gamma}}{\cosh\gamma}
 =\tanh\abs{\gamma}.
 \label{eq:imperfectbellmapping}
\end{equation}
However, the leading magic-injection formulas assume a stabilizer input and a
large acted-on subsystem, $\abs{A}\gg1$ \cite{Hou2026Injection}.  At
intermediate $\gamma$, the present input is
already non-stabilizer and has $\abs{A}=1$.  The many-copy, leading-order
imperfect-Bell-pair discussion of that work motivates a future local
non-Clifford quench, but its asymptotic formulas cannot be substituted into
the current two-qubit result.

Although the large-subsystem expression cannot be imported, the present
single-pair problem can be evaluated exactly without any asymptotic
approximation.  Apply a local unitary to subsystem $A$,
\begin{equation}
 \ket{\psi_{\gamma,U}}
 =(U_A\otimes I_B)\ket{\psi_\gamma},
 \qquad U_A\sim {\rm Haar}[U(2)],
 \label{eq:localHaarquench}
\end{equation}
and retain the associated linear stabilizer entropy
\cite{Hou2026Injection},
\begin{equation}
 Y^{\rm lin}(\psi)
 =1-\frac14\sum_{P\in\mathcal P_2}
 \langle\psi|P|\psi\rangle^4.
 \label{eq:Ylindefinition}
\end{equation}
For the sign-corrected state in Eq.~\eqref{eq:trianglepsi}, introduce
\begin{equation}
 w_\gamma=2a_\gamma b_\gamma
 =\abs{\tanh(2\gamma)}= \vert{}v_\gamma\vert{},
 \qquad
 u^2+w_\gamma^2=1.
 \label{eq:uwrelation}
\end{equation}
Its one-body Bloch vectors and two-body correlation matrix are
\begin{equation}
 \bm r_A=\bm r_B=(0,0,u),
 \qquad
 T=\operatorname{diag}(w_\gamma,-w_\gamma,1),
 \label{eq:initialBlochT}
\end{equation}
where $T_{ij}=\langle\sigma_i\otimes\sigma_j\rangle$. One-body Pauli expectations with respect to $\vert{}\psi_\gamma\rangle = a_\gamma\vert{}00\rangle + b_\gamma\vert{}11\rangle$ are:
$\langle X \otimes \mathbb{I} \rangle_{\text{init}} =\langle\mathbb{I} \otimes X \rangle_{\text{init}} = \langle Y \otimes \mathbb{I} \rangle_{\text{init}} =\langle\mathbb{I} \otimes Y \rangle_{\text{init}} = 0$,
$\langle Z \otimes \mathbb{I} \rangle_{\text{init}} = \langle \mathbb{I} \otimes Z \rangle_{\text{init}} = a_\gamma^2 - b_\gamma^2 = |u|=u$;
and two-body Pauli expectations are:
$\langle X \otimes X \rangle_{\text{init}} = 2a_{\gamma}b_{\gamma} = \vert{}\tanh(2\gamma)\vert{} = w_{\gamma}$,
$\langle Y \otimes Y \rangle_{\text{init}} = -2a_{\gamma}b_{\gamma} = -\vert{}\tanh(2\gamma)\vert{} = -w_{\gamma}$,
$\langle Z \otimes Z \rangle_{\text{init}} = a_{\gamma}^2 + b_{\gamma}^2 = 1$,
and all the off-diagonal terms, like $\langle X \otimes Y \rangle_{\text{init}}$, are zero.
Thus $\langle \sigma_k \otimes \sigma_j \rangle_{\text{init}} = \delta_{kj} t_j$.

The adjoint action of $U_A$ induces a rotation $R\in SO(3)$,
\begin{equation}
 U_A^\dagger\sigma_iU_A
 =\sum_{j=x,y,z}R_{ij}\sigma_j.
 \label{eq:SU2SO3}
\end{equation}
It follows directly that the post-quench expectation values are
\begin{align}
 \langle\sigma_i\otimes I\rangle_U
 & = \langle\psi_\gamma\vert{} (U_A^\dagger \sigma_i U_A \otimes I_B) \vert{}\psi_\gamma\rangle=uR_{iz},\\
 \langle I \otimes \sigma_i\rangle_U &= \langle\psi_\gamma\vert{} (U_A^\dagger I U_A \otimes \sigma_i) \vert{}\psi_\gamma\rangle = \langle\psi_\gamma\vert{} (I \otimes \sigma_i) \vert{}\psi_\gamma\rangle = \langle I \otimes \sigma_i\rangle_{\text{init}},\\
 \langle I\otimes Z\rangle_U&=u,\qquad
 \langle I\otimes X\rangle_U
 =\langle I\otimes Y\rangle_U=0,\notag\\
 \langle\sigma_i\otimes\sigma_j\rangle_U
 &=R_{ij}t_j,
 \qquad
 (t_x,t_y,t_z)=(w_\gamma,-w_\gamma,1).
 \label{eq:postquenchPauli}
\end{align}
For a Haar-distributed $U_A$, every column of $R$ ($\vec{R}_j = (R_{xj}, R_{yj}, R_{zj})^T$) is a uniformly distributed
unit vector on the two-sphere.  Any fixed Cartesian component $x$ therefore
has density $1/2$ on $[-1,1]$, and hence
\begin{equation}
 \mathbb E_U[R_{ij}^4]
 =\frac12\int_{-1}^{1}x^4\,dx
 =\frac15.
 \label{eq:rotationfourthmoment}
\end{equation}
We may now average all sixteen Pauli strings explicitly.  The identity
contributes one, the unchanged $B$-local component contributes $u^4$, the
three $A$-local components contribute $\sum_{i=x,y,z} (u R_{iz})^4 =3u^4/5$, and the nine two-body
components contribute $3(1+2w_\gamma^4)/5$: Since after rotation the two-body correlation matrix elements are $\langle \sigma_i \otimes \sigma_j \rangle_U = \langle \psi_\gamma \vert{} (U_A^\dagger \sigma_i U_A \otimes \sigma_j) \vert{} \psi_\gamma \rangle
= \sum_{k \in \{x,y,z\}} R_{ik} \langle \psi_\gamma \vert{} (\sigma_k \otimes \sigma_j) \vert{} \psi_\gamma \rangle = \sum_{k \in \{x,y,z\}} R_{ik} (\delta_{kj} t_j) = R_{ij} t_j$,
$i, j \in \{x, y, z\}$,
using $\mathbb{E}_U \left[ \sum_{i=x,y,z} R_{ij}^4 \right] = \mathbb{E}_U [R_{xj}^4] + \mathbb{E}_U [R_{yj}^4] + \mathbb{E}_U [R_{zj}^4] =\frac{3}{5}$, $\sum_{j=x,y,z} t_j^4 = w_{\gamma}^4 + (-w_{\gamma})^4 + 1^4 = 1 + 2w_{\gamma}^4$, we obtain the Haar average of two-body correlation $\mathbb{E}_U \left[ \sum_{i,j} (R_{ij} t_j)^4 \right] = \left( \mathbb{E}_U \sum_{i} R_{ij}^4 \right) \cdot \sum_{j} t_j^4 = \frac{3}{5} \left( w_{\gamma}^4 + (-w_{\gamma})^4 + 1^4 \right) = \frac{3}{5} (1 + 2w_{\gamma}^4)$.  Therefore
\begin{align}
 \mathbb E_U\sum_{P\in\mathcal P_2}\langle P\rangle_U^4
 &=1+u^4+\frac35u^4
 +\frac35(1+2w_\gamma^4)\notag\\
 &=1+u^4+\frac35u^4
 +\frac35\left[1+2(1-u^2)^2\right]\notag\\
 &=\frac{14-12u^2+14u^4}{5}.
 \label{eq:Haarquenchfourthsum}
\end{align}
Substitution into Eq.~\eqref{eq:Ylindefinition} yields the exact finite-size
result of the averaged linear stabilizer entropy after quench
\begin{equation}
 \overline{Y^{\rm lin}}(\gamma)
 :=\mathbb E_{U_A}Y^{\rm lin}(\psi_{\gamma,U})
 =\frac{3+6u^2-7u^4}{10},
 \qquad u=\operatorname{sech}(2\gamma).
 \label{eq:HaarquenchYlin}
\end{equation}
This expression has several useful checks.  At the product-stabilizer endpoint
$\gamma=0$, $\overline{Y^{\rm lin}}(0)=1/5$, which is the mean magic generated
by a one-qubit Haar unitary acting on a product stabilizer.  At the maximally
entangled Bell endpoint,
$\lim_{\abs{\gamma}\to\infty}\overline{Y^{\rm lin}}(\gamma)=3/10$, showing the
finite-size enhancement between the two stabilizer endpoints.
The full function is nonmonotonic: differentiating with respect to
$x=u^2$ gives $d\overline{Y^{\rm lin}}/dx=(6-14x)/10$,
so its maximum is $\overline{Y^{\rm lin}}_{\max}=3/7$ at $u^2=3/7$.

The preexisting linear magic of the unquenched state is obtained from
Eq.~\eqref{eq:paulifourthsum} as
$Y_{\rm init}^{\rm lin}=1-(1-u^2+u^4)=u^2(1-u^2)$.
Note that $\mathcal{M}_2^{\rm Sch}= -\log_2(1 - Y^{\rm lin}_{\rm init})$.
The increase in the linear quantity is consequently
\begin{align}
 \Delta\overline{Y^{\rm lin}}
 &:=\overline{Y^{\rm lin}}-Y_{\rm init}^{\rm lin}\notag\\
 &=\frac{3-4u^2+3u^4}{10}.
 \label{eq:HaarquenchDeltaY}
\end{align}
It reaches its minimum $1/6$ at $u^2=2/3$.  Thus the endpoint comparison
$1/5\to3/10$ confirms enhancement by stabilizer entanglement, while the
intermediate non-stabilizer states retain a nontrivial initial-magic
dependence.  This is precisely why the many-copy imperfect-Bell trend should
not be identified with a monotonic theorem for one pair.

Finally, one may define the annealed second stabilizer R\'enyi quantity
\begin{align}
 \widetilde M_{2,\rm ann}
 &:=-\log_2\left[1-\overline{Y^{\rm lin}}\right]\notag\\
 &=-\log_2\left[
 \frac{7-6u^2+7u^4}{10}
 \right].
 \label{eq:HaarquenchAnnealed}
\end{align}
Because the logarithm and Haar average do not commute,
$\mathbb E_U[M_2(\psi_{\gamma,U})]\ne\widetilde M_{2,\rm ann}$ in general.
Equation~\eqref{eq:HaarquenchYlin}, rather than
Eq.~\eqref{eq:HaarquenchAnnealed}, is the exact averaged quantity derived
above.

\paragraph{Replica and projected-ensemble moments.}

The present zero-dimensional reduced state also admits an algebraic
specialization of a twisted R\'enyi-$N$ correlator
\cite{Sala2026Holography}.  To avoid confusion between the replica number and the subsystem
dimension $\mathcal N$, denote the positive replica index by $\ell$.  Choosing
the Schmidt-level exchange operator $X$ gives
\begin{equation}
 C_X(\ell)
 :=\frac{\Tr\!\left[
 (\rho_A^R)^\ell X(\rho_A^R)^\ell X
 \right]}
 {\Tr[(\rho_A^R)^{2\ell}]},
 \label{eq:twistedRenyiDefinition}
\end{equation}
where $\text{Tr}[(\rho_A^R)^2] = p_+^2 + p_-^2 = (p_+ + p_-)^2 - 2p_+ p_-
= 1^2 - 2\left(\frac{C_\gamma^2}{4}\right) = 1 - \frac{C_\gamma^2}{2} $ for reduced density matrix $\rho_A^R = p_+ \vert{}0\rangle\langle 0\vert{} + p_- \vert{}1\rangle\langle 1\vert{}$ with Schmidt probability $p_+ = a_\gamma^2, p_- = b_\gamma^2$ and $p_+ + p_- = a_\gamma^2 + b_\gamma^2 = 1$, $p_+ p_- = a_\gamma^2 b_\gamma^2 = \frac{C_\gamma^2}{4}$.
From Eq.~\eqref{eq:schmidt},
$(\rho_A^R)^\ell=\operatorname{diag}(p_+^\ell,p_-^\ell)$.
one obtains step by step
\begin{align}
 (\rho_A^R)^\ell X
 &=
 \begin{pmatrix}
  0&p_+^\ell\\
  p_-^\ell&0
 \end{pmatrix},\notag\\
 (\rho_A^R)^\ell X(\rho_A^R)^\ell X
 &=(p_+p_-)^\ell I,
 \label{eq:twistedRenyiMatrixProduct}\\
 \Tr[(\rho_A^R)^{2\ell}]
 &=p_+^{2\ell}+p_-^{2\ell}.
\end{align}
It follows that
\begin{align}
 C_X(\ell)
 &=\frac{2(p_+p_-)^\ell}
 {p_+^{2\ell}+p_-^{2\ell}}\notag\\
 &=\frac{2(1-u^2)^\ell}
 {(1+u)^{2\ell}+(1-u)^{2\ell}}.
 \label{eq:twistedRenyiClosed}
\end{align}
Introducing the entanglement-level gap
\begin{equation}
 \Delta\epsilon
 :=\ln\frac{p_+}{p_-}= \ln(\coth^2\gamma),
 \label{eq:entanglementgap}
\end{equation}
the same result takes the compact form
\begin{equation}
 C_X(\ell)=\operatorname{sech}(\ell\Delta\epsilon)= \frac{2}{e^{l\Delta\epsilon} + e^{-l\Delta\epsilon}}= \frac{1}{\cosh(l\Delta\epsilon)} .
 \label{eq:twistedRenyiGap}
\end{equation}
Thus $p_+ = \frac{e^{\Delta\epsilon}}{1 + e^{\Delta\epsilon}} = \frac{1}{1 + e^{-\Delta\epsilon}} = \frac{e^{\Delta\epsilon/2}}{2\cosh(\Delta\epsilon/2)}= \frac{1}{2} \left[ 1 + \frac{2}{e^{2\gamma} + e^{-2\gamma}} \right] = \frac{(e^\gamma + e^{-\gamma})^2}{2(e^{2\gamma} + e^{-2\gamma})} = \frac{\cosh^2\gamma}{\cosh(2\gamma)}$,
$p_- = \frac{1}{1 + e^{\Delta\epsilon}} = \frac{e^{-\Delta\epsilon}}{1 + e^{-\Delta\epsilon}} = \frac{e^{-\Delta\epsilon/2}}{2\cosh(\Delta\epsilon/2)} = \frac{1}{2} \left[ 1 - \frac{2}{e^{2\gamma} + e^{-2\gamma}} \right] = \frac{(e^\gamma - e^{-\gamma})^2}{2(e^{2\gamma} + e^{-2\gamma})} = \frac{\sinh^2\gamma}{\cosh(2\gamma)}$.
Note that $p_\pm =  \frac{1 \pm u}{2} $ for $u= \operatorname{sech}(2\gamma) \in (0, 1]$. 
For finite $l$, the separable state ($\gamma=0$) has $u=\operatorname{sech}(0)=1$, $p_+=1$, $p_-=0$, $\Delta\epsilon\to\infty$, and $C_X(l)=\operatorname{sech}(\infty)=0$.  The maximally entangled state ($\vert{}\gamma\vert{}\to\infty$) instead has $u=\operatorname{sech}(\infty)=0$, $p_\pm=1/2$, $\Delta\epsilon=\ln(1)=0$, and $C_X(l)=\operatorname{sech}(0)=1$.  For $l=1$, $C_X(l)$ reduces to $\frac{1-u^2}{1+u^2}$.  In the R\'enyi-zero limit $l\to0$, $C_X(0)=\frac{\operatorname{Tr}[(\rho_A^0X)^2]}{\operatorname{Tr}[\rho_A^0]}=\frac{\operatorname{Tr}[X^2]}{\operatorname{Tr}[I_2]}=1$.
For every finite $\abs{\gamma}$ one has $p_+>p_-$ and hence
$C_X(\ell)\approx 2e^{-l\Delta\epsilon} = 2(\tanh^2\gamma)^l\to0$ as $\ell\to\infty$.
As long as $p_+ > p_-$, $\Delta\epsilon = \ln(p_+/p_-) > 0$, the entanglement spectrum possesses a strictly positive Schmidt gap. Consequently, as the replica index $\ell$ increases, the dominant term $p_+^{2\ell}$ in the denominator overwhelmingly suppresses the contribution of the smaller eigenvalue $p_-^{2\ell}$. This causes the cross-term $(p_+p_-)^\ell$ in the numerator to exhibit an exponential collapse relative to the denominator, directly yielding the asymptotic exponential decay behavior $C_X(\ell) = \frac{2e^{-\ell\Delta\epsilon}}{1 + e^{-2\ell\Delta\epsilon}} = 2e^{-\ell\Delta\epsilon} \left( 1 - e^{-2\ell\Delta\epsilon} + e^{-4\ell\Delta\epsilon} - \dots \right)\approx 2e^{-\ell\Delta\epsilon}$ in the large-$\ell$ limit, and $\operatorname{Tr}[\rho_A^{2\ell}] = p_+^{2\ell} + p_-^{2\ell} = p_+^{2\ell} \left[ 1 + \left(\frac{p_-}{p_+}\right)^{2\ell} \right] = p_+^{2\ell} \left[ 1 + e^{-2\ell\Delta\epsilon} \right]\approx p_+^{2\ell}$. The exponential decay here follows from the two-level Schmidt spectrum; the extended-system entanglement-holography and replica-order setting of Ref.~\cite{Sala2026Holography} involves a spatial bulk and a thermodynamic limit.
 This is a finite-dimensional nonlinear
repackaging of the Schmidt spectrum.  

The asymptotic decay of twisted correlator is related to effective ground-state projection. Rényi index $\ell$ acts as an effective inverse temperature (or imaginary-time length) for $-\ln \rho_A^R$, the large-$\ell$ limit acts as a zero-temperature cooling process that projects the normalized operator $\rho_A^\ell / \operatorname{Tr}[\rho_A^\ell]$ onto the non-degenerate ground subspace of  $-\ln \rho_A^R$ spanned by $\vert{}0\rangle_A$ (since $p_+ > p_-$ for any finite $\gamma$). Because the twist operator $\sigma_x$ is off-diagonal in the Schmidt basis ($\langle 0\vert{}_A \sigma_x \vert{}0\rangle_A = 0$), the twisted transition loop cannot be closed within this one-dimensional ground subspace alone. Completing the trace requires virtual excursions into the excited Schmidt state $\vert{}1\rangle_A$, which incurs an exponential suppression factor $(p_-/p_+)^\ell = e^{-\ell\Delta\epsilon}$, leading to $C_X(\ell) \sim 2e^{-\ell\Delta\epsilon} \to 0$. While in maximally entangled Bell limit ($\gamma \to \infty$, $\Delta\epsilon = 0$), the ground space of $-\ln \rho_A^R$ becomes two-fold degenerate ($p_+ = p_-$), preserving the full operator-loop amplitude $C_X(\ell) = \operatorname{sech}(0) = 1$ for all $\ell$.

Likewise, Hamiltonian state-design generation concerns unitary evolution
$e^{-iHt}$, assumes a $k$th-order no-resonance condition, and obtains its
principal analytic result after a GUE average and in the thermodynamic limit
\cite{Hou2026KDesign}.  The present
$H_\gamma$ is non-Hermitian and contains an exact repeated eigenvalue by
construction, so the no-resonance assumption is violated.  It would therefore
be incorrect to infer that the current isospectral family generates a state
$k$-design.

In the Schmidt basis, the normalized two-qubit right eigenstate reads $\vert{}\widetilde{R}_{00}\rangle = a_\gamma \vert{}0\rangle_A \vert{}0\rangle_B + b_\gamma \vert{}1\rangle_A \vert{}1\rangle_B$ with $a_\gamma^2 + b_\gamma^2 = 1$. Treating qubit $B$ as an auxiliary/environmental system, performing a projective measurement on $B$ in the Schmidt basis $\{\vert{}0\rangle_B, \vert{}1\rangle_B\}$ 
Schmidt basis (which, by Eq.\ref{eq:R00normalized}, is aligned with the computational basis) yields outcome $\vert{}0\rangle_B$ with probability $p_0 = \vert{}\langle 0_B \vert{} \widetilde{R}_{00}\rangle\vert{}^2 = a_\gamma^2$ (collapsing subsystem $A$ to $\vert{}0\rangle_A$), and outcome $\vert{}1\rangle_B$ with probability $p_1 = \vert{}\langle 1_B \vert{} \widetilde{R}_{00}\rangle\vert{}^2 = b_\gamma^2$ (collapsing subsystem $A$ to $\vert{}1\rangle_A$). This measurement generates a conditional discrete projected ensemble on subsystem $A$: $\left\{ (a_\gamma^2, \vert{}0\rangle_A), \; (b_\gamma^2, \vert{}1\rangle_A) \right\}$.
Since the two conditional states are
orthogonal, its $k$th frame potential is independent of $k\geq1$:
\begin{align}
 F_\gamma^{(k)}
 &:=\sum_{i,j}p_i p_j\abs{\braket{\psi_i}{\psi_j}}^{2k}
 =a_\gamma^4+b_\gamma^4
 \notag\\
 &=\Tr[(\rho_A^R)^2]
 =e^{-S_A^{(2,R)}}
 =\frac{1+u^2}{2}, \label{eq:ProjectedEnsembleFramePotential}
\end{align}
where $p_+=q_{00}=a_{\gamma}^2= \frac{\cosh^2\gamma}{\cosh(2\gamma)}$, $p_-=q_{11}=b_{\gamma}^2= \frac{\sinh^2\gamma}{\cosh(2\gamma)}$.
Thus the projected-ensemble frame potential, the reduced purity, and the
Schmidt-basis IPR coincide in this model.  For a Haar-random pure qubit the
exact value is
\begin{equation}
 F_{\rm Haar}^{(k)}= \int_{\mathrm{Haar}} d\psi \int_{\mathrm{Haar}} d\phi \, \vert{}\langle\psi\vert{}\phi\rangle\vert{}^{2k} = \frac{k!}{\prod_{m=0}^{k-1}(d_A + m)} = \frac{k! \cdot (d_A - 1)!}{(d_A + k - 1)!}=\frac{1}{k+1},
 \label{eq:QubitHaarFramePotential}
\end{equation}
where $d_A = 2$.
Equation~\eqref{eq:ProjectedEnsembleFramePotential} reaches the Haar first
moment only in the Bell limit, where $F_\gamma^{(1)}=1/2$.  Even there the
fixed Schmidt-basis measurement gives $F_\gamma^{(k)}=1/2>1/(k+1)$ for every
$k\geq2$, so the two-state projected ensemble is a state 1-design but not a
2-design.  For finite $\gamma$, its first moment is
$\rho_A^R\neq\mathbb I_2/2$, and it is not even a 1-design.  Tensor-network
confinement requires extended random-MPS or shallow-circuit geometries; only
the projected-ensemble and frame-potential diagnostics are used here
\cite{Lami2026Confinement}.

\subsection{Bell nonlocality from the correlation tensor}

The same Schmidt data also determine the maximal CHSH violation.  This gives
an operational Bell-nonlocality diagnostic distinct from the Bell-operator
level statistics used in many-body studies.  We use here the standard
positive right-state density operator
\begin{equation}
 \rho_\gamma^R
 =\ket{\widetilde R_{00}}\bra{\widetilde R_{00}},
 \qquad
 \ket{\widetilde R_{00}}=a\ket{00}+b\ket{11},
 \label{eq:CHSHRightState}
\end{equation}
with the conventional Born rule.  Define its real $3\times3$ correlation
matrix by
\begin{equation}
 (T_\gamma)_{ij}
 :=\Tr\!\left[\rho_\gamma^R
 (\sigma_i\otimes\sigma_j)\right],
 \qquad i,j\in\{x,y,z\}.
 \label{eq:CHSHCorrelationDefinition}
\end{equation}
The diagonal entries can be evaluated directly.  Since
\begin{align}
 (X\otimes X)\ket{00}&=\ket{11},
 & (X\otimes X)\ket{11}&=\ket{00},\notag\\
 (Y\otimes Y)\ket{00}&=-\ket{11},
 & (Y\otimes Y)\ket{11}&=-\ket{00},\notag\\
 (Z\otimes Z)\ket{00}&=\ket{00},
 & (Z\otimes Z)\ket{11}&=\ket{11},
 \label{eq:CHSHPauliActions}
\end{align}
one obtains
\begin{align}
 \langle X\otimes X\rangle
 &=ab+ab=2ab=v_\gamma,\notag\\
 \langle Y\otimes Y\rangle
 &=-ab-ab=-2ab=-v_\gamma,\notag\\
 \langle Z\otimes Z\rangle
 &=a^2+b^2=1.
 \label{eq:CHSHDiagonalCorrelators}
\end{align}
Every off-diagonal correlator vanishes because the state has real Schmidt
coefficients and support only on $\ket{00}$ and $\ket{11}$.  Therefore
\begin{equation}
 T_\gamma
 =\operatorname{diag}(v_\gamma,-v_\gamma,1),
 \qquad
 v_\gamma=2ab=\tanh(2\gamma),
 \qquad C_\gamma=\abs{v_\gamma}.
 \label{eq:CHSHCorrelationMatrix}
\end{equation}
The Horodecki criterion states that the maximal CHSH value of a two-qubit
state is $2\sqrt{m_1+m_2}$, where $m_1$ and $m_2$ are the two largest
eigenvalues of $T_\gamma^{\mathsf T}T_\gamma$~\cite{Horodecki1995}.  In the
present case,
\begin{equation}
 T_\gamma^{\mathsf T}T_\gamma
 =\operatorname{diag}(v_\gamma^2,v_\gamma^2,1)
 =\operatorname{diag}(C_\gamma^2,C_\gamma^2,1),
 \label{eq:CHSHCorrelationSingularValues}
\end{equation}
and $0\leq C_\gamma\leq1$, so the two largest eigenvalues are $1$ and
$C_\gamma^2$.  It follows that
\begin{align}
 \mathcal B_{\max}(\gamma)
 &=2\sqrt{1+C_\gamma^2}\notag\\
 &=2\sqrt{1+\tanh^2(2\abs\gamma)}.
 \label{eq:CHSHMaximum}
\end{align}
At the Hermitian point, $C_0=0$ and
$\mathcal B_{\max}(0)=2$, which only saturates the local-hidden-variable
bound.  For every finite $\gamma\neq0$, $\mathcal B_{\max}(\gamma)>2$, so the
normalized right eigenstate violates the CHSH inequality.  In the maximally
entangled limit,
$\lim_{\abs\gamma\rightarrow\infty}\mathcal B_{\max}(\gamma)=2\sqrt2$, which
is the Tsirelson bound.

Equations~\eqref{eq:entropyConcurrenceDictionary} and
\eqref{eq:CHSHMaximum} give the direct entropy--nonlocality relation
$\mathcal B_{\max}=2\sqrt{3-2e^{-S_A^{(2,R)}}}$.  Likewise, because the
pure-state PPT negativity is $\mathcal N_{\rm PPT}=C_\gamma/2$, one has
$\mathcal B_{\max}=2\sqrt{1+4\mathcal N_{\rm PPT}^2}$.
Thus CHSH nonlocality is monotonic in the entanglement of this pure
two-qubit family and does not supply an independent state variable.  Its
usefulness is operational: it converts the Schmidt entanglement into a Bell
test.  This differs sharply from the nonmonotonic exact non-local magic in
Eq.~\eqref{eq:magicgamma}.  The derivation should not be transferred to the
biorthogonal reduction, whose non-positive or complex weights need not define
physical measurement probabilities.

\subsection{Open-system preparation and decoherence}

\paragraph{Dissipative stabilization.}

The Schmidt family in Eq.~\eqref{eq:trianglepsi} also has a direct open-system
embedding.  This does not change the static Hamiltonian construction above;
rather, it supplies an operational dissipative mechanism that prepares the
same normalized right state.  Using the absolute value of the original
non-Hermiticity parameter directly, write
\begin{equation}
 \ket{\psi_\gamma}
 =\frac{\cosh(\abs{\gamma})\ket{00}
 +\sinh(\abs{\gamma})\ket{11}}
 {\sqrt{\cosh(2\abs{\gamma})}}.
 \label{eq:dissipativeSchmidtState}
\end{equation}
For $\gamma<0$, Eq.~\eqref{eq:dissipativeSchmidtState} is related to the
original right eigenvector by the local Clifford operation already used below
Eq.~\eqref{eq:magicbound}.  Introduce the two jump operators
\begin{align}
 J_A(\abs{\gamma})
 &=\cosh(\abs{\gamma})\,\sigma_A^-
   -\sinh(\abs{\gamma})\,\sigma_B^+,
 \notag\\
 J_B(\abs{\gamma})
 &=\cosh(\abs{\gamma})\,\sigma_B^-
   -\sinh(\abs{\gamma})\,\sigma_A^+,
 \label{eq:TMSjumpoperators}
\end{align}
and the two-mode-squeezing Lindbladian
\begin{equation}
 \mathcal L_\gamma(\rho)
 =\kappa\mathcal D[J_A(\abs{\gamma})]\rho
 +\kappa\mathcal D[J_B(\abs{\gamma})]\rho,
 \qquad
 \mathcal D[J]\rho
 =J\rho J^\dagger-\frac12\{J^\dagger J,\rho\},
 \label{eq:TMSLindbladian}
\end{equation}
where $\kappa>0$ is a dissipative rate and is distinct from the condition
number $\kappa_2(S_\gamma)$ in Eq.~\eqref{eq:kappa}.  The dark-state property
can be verified without solving the master equation.  Using
\begin{align}
 \sigma_A^-\ket{11}=\ket{01},
 &\qquad \sigma_B^+\ket{00}=\ket{01},
 \notag\\
 \sigma_B^-\ket{11}=\ket{10},
 &\qquad \sigma_A^+\ket{00}=\ket{10},
\end{align}
where $\sigma_A^- = \vert{}0\rangle_A \langle1\vert_A \otimes \mathbb{I}_B$, $\sigma_B^+ = \mathbb{I}_A \otimes \vert{}1\rangle_B \langle0\vert_B$,
$\sigma_A^+ = \vert{}1\rangle_A \langle0\vert_A \otimes \mathbb{I}_B$, $\sigma_B^-= \mathbb{I}_A \otimes \vert{}0\rangle_B \langle1\vert_B$,
$\sigma_A^+ \vert{}00\rangle = (\sigma_A^+ \vert{}0\rangle_A) \otimes \vert{}0\rangle_B = \vert{}1\rangle_A \otimes \vert{}0\rangle_B = \vert{}10\rangle$,
$\sigma_B^- \vert{}00\rangle = \vert{}0\rangle_A \otimes (\sigma_B^- \vert{}0\rangle_B) =0$,
$\sigma_A^+ \vert{}11\rangle = (\sigma_A^+ \vert{}1\rangle_A) \otimes \vert{}1\rangle_B = 0$,
$\sigma_B^- \vert{}11\rangle = \vert{}1\rangle_A \otimes (\sigma_B^- \vert{}1\rangle_B) = \vert{}1\rangle_A \otimes \vert{}0\rangle_B = \vert{}10\rangle$,
one obtains
\begin{align}
 J_A\ket{\psi_\gamma}
 &=\frac{
 \cosh(\abs{\gamma})\sinh(\abs{\gamma})
 -\sinh(\abs{\gamma})\cosh(\abs{\gamma})}
 {\sqrt{\cosh(2\abs{\gamma})}}\ket{01}=0,
 \notag\\
 J_B\ket{\psi_\gamma}
 &=\frac{
 \cosh(\abs{\gamma})\sinh(\abs{\gamma})
 -\sinh(\abs{\gamma})\cosh(\abs{\gamma})}
 {\sqrt{\cosh(2\abs{\gamma})}}\ket{10}=0.
 \label{eq:TMSdarkstatecheck}
\end{align}
Therefore each dissipator in Eq.~\eqref{eq:TMSLindbladian} annihilates
$\rho_\gamma=\ket{\psi_\gamma}\bra{\psi_\gamma}$ and
$\mathcal L_\gamma(\rho_\gamma)=0$.
This is the same Schmidt-diagonal dark state used as the communication-qubit
resource in Floquet reservoir engineering \cite{Yao2026Floquet}.

The corresponding entropy--concurrence identity is collected in
Appendix~\ref{app:EntropyConcurrence}.

This embedding is not a dynamical property of $H_\gamma$
itself.  The Floquet protocol requires additional logical qubits, communication
qubits, controlled local gates, and the Lindbladian in
Eq.~\eqref{eq:TMSLindbladian}.  Its role here is to show that the same Schmidt
family generated by the non-unitary similarity map can serve as an exactly
matched dissipative entanglement resource.

\paragraph{Local dephasing and resource decay.}

The charge-dephasing channel used to probe extended SPT systems can be applied
exactly to the present two-qubit state without importing the spatial SPT or
strong-to-weak symmetry-breaking conclusions of that setting
\cite{Lu2026Dephasing}.  Acting only on subsystem $B$, define
\begin{equation}
 \mathcal E_p^{(B)}(\rho)
 =(1-2p)\rho
 +2p\sum_{j=0}^{1}
 (I_A\otimes|j\rangle\langle j|)\rho
 (I_A\otimes|j\rangle\langle j|),
 \qquad 0\leq p\leq\frac12,
 \label{eq:LocalChargeDephasingChannel}
\end{equation}
and introduce the surviving coherence factor $\eta:=1-2p$.
At $p=0$, the initial state is $\rho_\gamma(0) = \vert{}\psi_\gamma\rangle\langle\psi_\gamma\vert{}= (a_\gamma\vert{}00\rangle + b_\gamma\vert{}11\rangle) (a_\gamma\langle00\vert{} + b_\gamma\langle11\vert{}) = a_\gamma^2\vert{}00\rangle\langle00\vert{} + a_\gamma b_\gamma\vert{}00\rangle\langle11\vert{} + a_\gamma b_\gamma\vert{}11\rangle\langle00\vert{} + b_\gamma^2\vert{}11\rangle\langle11\vert{}$
, we have
\begin{equation}\begin{aligned}
&\sum_{j=0}^1 (I_A \otimes \vert{}j\rangle\langle j\vert{}) \left( a_\gamma^2 \vert{}0\rangle\langle0\vert{}_A \otimes \vert{}0\rangle\langle0\vert{}_B \right) (I_A \otimes \vert{}j\rangle\langle j\vert{})
= a_\gamma^2 \vert{}0\rangle\langle0\vert{}_A \otimes \left( \sum_{j=0}^1 \vert{}j\rangle\langle j\vert{}0\rangle\langle0\vert{}j\rangle\langle j\vert{} \right)= a_\gamma^2 \vert{}00\rangle\langle00\vert{},\\
&\sum_{j=0}^1 (I_A \otimes \vert{}j\rangle\langle j\vert{}) \left( b_\gamma^2 \vert{}1\rangle\langle 1\vert{}_A \otimes \vert{}1\rangle\langle 1\vert{}_B \right) (I_A \otimes \vert{}j\rangle\langle j\vert{})
= b_\gamma^2 \vert{}11\rangle\langle11\vert{},\\
&\sum_{j=0}^1 (I_A \otimes \vert{}j\rangle\langle j\vert{}) \left( a_\gamma b_\gamma \vert{}0\rangle\langle1\vert{}_A \otimes \vert{}0\rangle\langle1\vert{}_B \right) (I_A \otimes \vert{}j\rangle\langle j\vert{})
= a_\gamma b_\gamma \vert{}0\rangle\langle1\vert{}_A \otimes \left( \sum_{j=0}^1 \vert{}j\rangle\langle j\vert{}0\rangle\langle 1\vert{}j\rangle\langle j\vert{} \right)=0.
\end{aligned}\end{equation}
Thus
\begin{equation}\begin{aligned}
\mathcal{E}_p^{(B)}(\rho_\gamma)
&=(1-2p)\Big[a_\gamma^2|00\rangle\langle00|
 +b_\gamma^2|11\rangle\langle11|\\
&\qquad\quad+a_\gamma b_\gamma|00\rangle\langle11|
 +a_\gamma b_\gamma|11\rangle\langle00|\Big]
 +2p\Big[a_\gamma^2|00\rangle\langle00|
 +b_\gamma^2|11\rangle\langle11|\Big]\\
&=a_\gamma^2|00\rangle\langle00|+b_\gamma^2|11\rangle\langle11|
 +\eta a_\gamma b_\gamma|00\rangle\langle11|
 +\eta a_\gamma b_\gamma|11\rangle\langle00|.
\end{aligned}\end{equation}
and the output in the
ordered basis $\{|00\rangle,|01\rangle,|10\rangle,|11\rangle\}$ is
\begin{equation}
 \rho_\gamma(p)
 =\mathcal E_p^{(B)}(\rho_\gamma)
 =\begin{pmatrix}
 a_\gamma^2&0&0&\eta a_\gamma b_\gamma\\
 0&0&0&0\\
 0&0&0&0\\
 \eta a_\gamma b_\gamma&0&0&b_\gamma^2
 \end{pmatrix}.
 \label{eq:DephasedSchmidtDensityMatrix}
\end{equation}
Writing $C_\gamma=2a_\gamma b_\gamma=|\tanh(2\gamma)|$, its global purity is
\begin{align}
 \Tr[\rho_\gamma(p)^2]
 &=a_\gamma^4+b_\gamma^4
 +2\eta^2a_\gamma^2b_\gamma^2\notag\\
 &=1-\frac{1-\eta^2}{2}C_\gamma^2.
 \label{eq:DephasedGlobalPurity}
\end{align}
Since $(|i_A,j_B\rangle\langle k_A,l_B|)^{T_B}
=|i_A,l_B\rangle\langle k_A,j_B|$, the off-diagonal terms are mapped as
$(|00\rangle\langle11|)^{T_B}=|01\rangle\langle10|$ and
$(|11\rangle\langle00|)^{T_B}=|10\rangle\langle01|$.  Therefore
\begin{equation*}\begin{aligned}
\rho_\gamma(p)^{T_B}
&=a_\gamma^2|00\rangle\langle00|+b_\gamma^2|11\rangle\langle11|
 +\eta a_\gamma b_\gamma|01\rangle\langle10|
 +\eta a_\gamma b_\gamma|10\rangle\langle01|\\
&=\begin{pmatrix}
a_\gamma^2&0&0&0\\
0&0&\eta a_\gamma b_\gamma&0\\
0&\eta a_\gamma b_\gamma&0&0\\
0&0&0&b_\gamma^2
\end{pmatrix}.
\end{aligned}\end{equation*}
Partial transposition with respect to $B$ produces the spectrum
\begin{equation}
 \operatorname{Spec}[\rho_\gamma(p)^{T_B}]
 =\left\{a_\gamma^2,b_\gamma^2,
 \frac{\eta C_\gamma}{2},-\frac{\eta C_\gamma}{2}\right\},
 \label{eq:DephasedPartialTransposeSpectrum}
\end{equation}
so the PPT negativity and concurrence decay linearly with the surviving
coherence,
\begin{equation}
 \mathcal N_{\rm PPT}(p)=\frac{\eta C_\gamma}{2},
 \qquad
 C[\rho_\gamma(p)]=\eta C_\gamma.
 \label{eq:DephasedNegativityConcurrence}
\end{equation}
The two-qubit correlation matrix is
\begin{equation}
 T_\gamma(p)
 =\operatorname{diag}(\eta C_\gamma,-\eta C_\gamma,1).
 \label{eq:DephasedCorrelationMatrix}
\end{equation}
The Horodecki criterion then gives
\begin{equation}
 \mathcal B_{\max}(p)
 =2\sqrt{1+\eta^2C_\gamma^2}
 =2\sqrt{1+4\mathcal N_{\rm PPT}(p)^2}.
 \label{eq:DephasedCHSHMaximum}
\end{equation}
For every entangled member of the family and every $p<1/2$, a nonzero
coherence remains and Eq.~\eqref{eq:DephasedCHSHMaximum} is larger than two.
At complete dephasing, $p=1/2$ and $\eta=0$, the state becomes
$\rho_\gamma(1/2)=a_\gamma^2|00\rangle\langle00|
+b_\gamma^2|11\rangle\langle11|$, with
$\mathcal N_{\rm PPT}(1/2)=0$ and $\mathcal B_{\max}(1/2)=2$.
The output is then a separable but classically correlated mixture.  In
contrast, trace preservation on $B$ implies
\begin{equation}
 \rho_A(p)=\Tr_B[\rho_\gamma(p)]
 =a_\gamma^2|0\rangle\langle0|
 +b_\gamma^2|1\rangle\langle1|
 =\rho_A(0),
 \label{eq:ReducedStateInvariantUnderDephasing}
\end{equation}
so $S_A^{(\mathrm{vN})}$ and $S_A^{(2)}$ remain unchanged throughout the
channel.  This explicitly demonstrates why the entropy of one subsystem is an
entanglement measure only when the global state is pure: after dephasing, the
same local entropy coexists with continuously decreasing negativity and CHSH
violation, and at $p=1/2$ it contains only classical correlation and local
mixedness.  Extended-system SPT, string-order, and SWSSB conclusions require
a spatial bulk, protecting symmetries, and a thermodynamic limit, none of
which is present in the current model \cite{Lu2026Dephasing}.

\paragraph{Model-constrained reconstruction.}

Within the normalized pure-state manifold, the single marginal observable
$u=\langle Z\otimes I\rangle$ fixes
$\rho_A^R=(I+uZ)/2$ and therefore reconstructs the deformation magnitude as
$|\gamma|=\tfrac12\operatorname{arcosh}(1/u)$.  No additional state variable
is needed: the same measurement gives
$C_\gamma=\sqrt{1-u^2}$,
$\mathcal N_{\rm PPT}=\sqrt{1-u^2}/2$,
$\mathcal B_{\max}=2\sqrt{2-u^2}$, and
$\mathcal M_2^{\rm NL}=-\log_2(1-u^2+u^4)$.  Thus a single local expectation
value reconstructs the entanglement, Bell-nonlocality, and non-local-magic
diagnostics only because purity and the one-parameter Schmidt form are known
in advance.

Equation~\eqref{eq:ReducedStateInvariantUnderDephasing} gives the sharp
counterexample once this purity assumption is removed: all values of the
dephasing strength have the same $\rho_A(p)$ and the same $u$.  The missing
coherence is recovered by one two-body correlator,
$\langle X\otimes X\rangle=\eta\sqrt{1-u^2}$, so that
$\eta=\langle X\otimes X\rangle/\sqrt{1-u^2}$ whenever $0\leq u<1$.
The pair $(u,\langle X\otimes X\rangle)$ then determines the complete matrix
in Eq.~\eqref{eq:DephasedSchmidtDensityMatrix}, including its negativity and
maximal CHSH value.  At the product endpoint $u=1$ the coherence vanishes and
the value of $\eta$ is operationally irrelevant.  This contrast separates
model-constrained pure-state reconstruction from mixed-state tomography
without introducing an external reconstruction algorithm.

\subsection{Magic preparation, concentration, and stabilizer distance}

\paragraph{Analytic preparation circuit.}

Equation~\eqref{eq:trianglecnot} also reduces preparation of the entangled
right state to one single-qubit rotation followed by a Clifford gate.  Since
the CNOT is self-inverse,
\begin{equation}
 |\psi_\gamma\rangle
 =\operatorname{CNOT}_{A\rightarrow B}
 \left[
 R_y^{(A)}(2\theta_\gamma)|0\rangle_A\otimes|0\rangle_B
 \right],
 \qquad
 \theta_\gamma=\arctan[\tanh(\abs\gamma)].
 \label{eq:FaultTolerantRyPreparation}
\end{equation}
Thus the complete two-qubit family is prepared exactly by one continuously
tunable single-qubit rotation followed by one CNOT.  No separate
fault-tolerant resource-state construction is required for the analytical
claims of this work.

\paragraph{Exact stabilizer fidelity.}

The Clifford reduction in Eq.~\eqref{eq:trianglecnot} also makes the distance
from the stabilizer set explicit.  Stabilizer fidelity is invariant under
Clifford transformations and is unchanged by adjoining the stabilizer ancilla
$\ket{0}$.  Therefore the two-qubit fidelity can be evaluated from
$\ket{\phi_\gamma}=a_\gamma\ket{0}+b_\gamma\ket{1}$.  The relevant maximal
overlaps are attained by a computational-basis stabilizer and an
$X$-basis stabilizer, or equivalently by $\ket{00}$ and
$\ket{\Phi^+}=(\ket{00}+\ket{11})/\sqrt{2}$ before the CNOT.  Hence
\begin{align}
 F_{\rm Stab}(\psi_\gamma)
 &:=\max_{\rho \in \mathcal{S}} \langle\psi_\gamma\vert{}\rho\vert{}\psi_\gamma\rangle =\max_{\ket{s}\in{\rm Stab}}\abs{\braket{s}{\psi_\gamma}}^2
 =\max\left\{a_\gamma^2,\frac{(a_\gamma+b_\gamma)^2}{2}\right\}
 \notag\\
 &=\frac{1+\max\{u,C_\gamma\}}{2},
 \qquad
 1-F_{\rm Stab}(\psi_\gamma)
 =\frac{1-\max\{u,C_\gamma\}}{2}.
 \label{eq:ExactStabilizerFidelity}
\end{align}
where $\rho = \sum_k p_k \vert{}s_k\rangle\langle s_k\vert{}$ ($p_k \ge 0, \sum p_k = 1$) and the free states (stabilizer mixed states) form the convex set $\mathcal{S} = \operatorname{conv}(\text{STAB})$, since
$$\langle\psi_\gamma\vert{}\rho\vert{}\psi_\gamma\rangle = \sum_k p_k \vert{}\langle s_k\vert{}\psi_\gamma\rangle\vert{}^2 \le \max_k \vert{}\langle s_k\vert{}\psi_\gamma\rangle\vert{}^2 \sum_k p_k = \max_{\vert{}s\rangle \in \text{STAB}} \vert{}\langle s\vert{}\psi_\gamma\rangle\vert{}^2$$.
Here for computational basis we have $\max_{s \in \{0, 1\}} \vert{}\langle s\vert{}\phi_\gamma\rangle\vert{}^2 = a_\gamma^2 = \frac{1+u}{2}$ since $\vert{}\langle 0\vert{}\phi_\gamma\rangle\vert{}^2 = a_\gamma^2$,
$\vert{}\langle 1\vert{}\phi_\gamma\rangle\vert{}^2 = b_\gamma^2$.
For X-basis, we have $\langle +\vert{}\phi_\gamma\rangle = \frac{\langle 0\vert{} + \langle 1\vert{}}{\sqrt{2}} (a_\gamma \vert{}0\rangle + b_\gamma \vert{}1\rangle) = \frac{a_\gamma + b_\gamma}{\sqrt{2}}$, $\vert{}\langle +\vert{}\phi_\gamma\rangle\vert{}^2 = \frac{1 + 2a_\gamma b_\gamma}{2} = \frac{1 + C_\gamma}{2}$, $\vert{}\langle -\vert{}\phi_\gamma\rangle\vert{}^2 = \frac{(a_\gamma - b_\gamma)^2}{2} = \frac{1 - 2a_\gamma b_\gamma}{2} = \frac{1 - C_\gamma}{2}$, and thus $\max_{s \in \{+, -\}} \vert{}\langle s\vert{}\phi_\gamma\rangle\vert{}^2 = \frac{(a_\gamma + b_\gamma)^2}{2} = \frac{1 + C_\gamma}{2}$.
Both endpoints are stabilizer states: $F_{\rm Stab}=1$ at $\gamma=0$ and in
the Bell limit $\abs\gamma\rightarrow\infty$ ($u=0$).  The fidelity is minimal when
$u=C_\gamma=1/\sqrt{2}$, namely
\begin{equation}
 \abs\gamma=\frac12\operatorname{arsinh}(1),
 \qquad
 F_{\rm Stab}^{\min}=\frac{1+1/\sqrt{2}}{2}.
 \label{eq:MinimumStabilizerFidelity}
\end{equation}
This is precisely the point where Eq.~\eqref{eq:magicgamma} is maximal, so
the exact non-local stabilizer R\'enyi entropy is maximal where the
stabilizer fidelity is minimal.  Bounded-memory stabilizer testing addresses
a distinct operational task: given copies of an unknown $n$-qubit pure state
and $k$ qubits of coherent memory, it tests whether the state is a stabilizer
state or has stabilizer fidelity at most $1-\epsilon$
\cite{Arunachalam2026StabilizerTesting}.  It does not directly estimate
$F_{\rm Stab}$ or the non-local stabilizer R\'enyi entropy.  For the present
known two-qubit family, Eq.~\eqref{eq:ExactStabilizerFidelity} already gives
the stabilizer fidelity exactly.

The spectral-geometry result also changes the interpretation of many copies.
For
\begin{equation}
 |\Psi_\gamma^{(m)}\rangle
 =|\psi_\gamma\rangle^{\otimes m},
 \qquad
 p=\frac{1+u}{2},\quad 1-p=\frac{1-u}{2},
 \label{eq:MulticopyDeformedDimers}
\end{equation}
the Schmidt probabilities are $p^{m-k}(1-p)^k$ with multiplicities
$\binom{m}{k}$.  Specializing the imperfect-dimer asymptotics to any fixed
$0<|\gamma|<\infty$ gives
\begin{equation}
 \mathcal M_2^{\rm NL}(\Psi_\gamma^{(m)})
 =\frac32\log_2 m+O_\gamma(1),
 \qquad m\rightarrow\infty,
 \label{eq:MulticopyNonlocalMagicScaling}
\end{equation}
where the optimization allows collective local unitaries within all $m$
qubits of $A$ and, independently, all $m$ qubits of $B$
\cite{Sierant2026SpectralGeometry}.  By contrast, retaining the original
pairwise Schmidt basis gives the additive stabilizer entropy
\begin{equation}
 M_2^{S}(|\psi_\gamma\rangle^{\otimes m})
 =m\left[-\log_2(1-u^2+u^4)\right].
 \label{eq:PairwiseBasisExtensiveSRE}
\end{equation}
Thus pairwise processing sees an extensive resource, whereas unrestricted
collective local basis changes reduce the true non-local contribution to
logarithmic growth.  The order of limits matters: the asymptotic statement
holds after fixing a finite nonzero $|\gamma|$; at the product endpoint
$\gamma=0$ and the Bell-pair endpoint $|\gamma|\to\infty$, the unrestricted
non-local magic is exactly zero for every finite $m$.

\paragraph{Universal multi-copy concentration.}

The Clifford reduction in Eq.~\eqref{eq:trianglecnot} isolates the pure
single-qubit state
\begin{equation}
 |\phi_\gamma\rangle
 =a_\gamma|0\rangle+b_\gamma|1\rangle.
 \label{eq:ConcentrationSingleQubitState}
\end{equation}
Since $x_\gamma = \langle \phi_\gamma \vert{} X \vert{} \phi_\gamma \rangle = 2a_\gamma b_\gamma = \frac{2\cosh\gamma\vert{}\sinh\gamma\vert{}}{\cosh(2\gamma)} = \vert{}\tanh(2\gamma)\vert{} = C_\gamma$, $y_\gamma = \langle \phi_\gamma \vert{} Y \vert{} \phi_\gamma \rangle = 0$, $z_\gamma = \langle \phi_\gamma \vert{} Z \vert{} \phi_\gamma \rangle = a_\gamma^2 - b_\gamma^2 = \frac{\cosh^2\gamma - \sinh^2\gamma}{\cosh(2\gamma)} = \operatorname{sech}(2\gamma) = u$,
its Bloch vector is
\begin{equation}
 (x_\gamma,y_\gamma,z_\gamma)
 =\left(2a_\gamma b_\gamma,0,a_\gamma^2-b_\gamma^2\right)
 =\left(C_\gamma,0,u\right),
 \qquad C_\gamma^2=1-u^2.
 \label{eq:ConcentrationBlochVector}
\end{equation}
The linearized order-three stabilizer R\'enyi entropy of a pure qubit is
\begin{equation}
 M_{3}^{\rm lin}(|\phi\rangle)
 =\frac12\left(1-x^6-y^6-z^6\right).
 \label{eq:LinearizedThirdSREDefinition}
\end{equation}
Consequently, the present one-parameter family obeys
\begin{align}
 M_{3}^{\rm lin}(\phi_\gamma)
 &=\frac12\left[1-(1-u^2)^3-u^6\right]\notag\\
 &=\frac32u^2(1-u^2).
 \label{eq:LinearizedThirdSREGamma}
\end{align}
The order-three specialization of
Eq.~\eqref{eq:TirritoGaussianSpectrumBound} is monotonically related to this
linearized quantity:
\begin{align}
 M_{3,>}^{\rm NL}
 &=-\frac12\log_2\!\left[
 \frac{1+u^6+(1-u^2)^3}{2}
 \right]\notag\\
 &=-\frac12\log_2\!\left[1-M_3^{\rm lin}(\phi_\gamma)\right].
 \label{eq:TirritoOrderThreeBridge}
\end{align}
This relation applies to the single-mode Gaussian Schmidt-spectrum weight.
For $\alpha>2$, $M_{3,>}^{\rm NL}$ denotes the Gaussian bound; unrestricted
local-unitary optimality remains open for general subsystem sizes.
A recent universal concentration protocol acts on copies of a
\emph{single-qubit} non-stabilizer state using stabilizer operations.
Its six-copy block is optimal within the protocol class studied in
Ref.~\cite{Rizzo2026Concentration}; the eight-copy construction achieves
the probability $2M_3^{\rm lin}/3$, for which global optimality has not
been established.  After applying the cross-partition Clifford decoding in
Eq.~\eqref{eq:trianglecnot}, the achievable
probabilities for the decoded states are
\begin{align}
 p_6(\gamma)
 &=\Pr\!\left(\phi_\gamma^{\otimes6}\rightarrow\mathrm{CCZ}\right)
 =\frac12u^2(1-u^2),
 \label{eq:SixCopyCCZProbability}\\
 p_8(\gamma)
 &=\Pr\!\left(\phi_\gamma^{\otimes8}\rightarrow\mathrm{CCZ}\right)
 =u^2(1-u^2).
 \label{eq:EightCopyCCZProbability}
\end{align}
These expressions vanish at both stabilizer endpoints, $\gamma=0$ and
$\abs\gamma\rightarrow\infty$, and are positive for every finite
$0<\abs\gamma<\infty$.  They are maximal when $u^2=1/2$, precisely where
Eq.~\eqref{eq:magicgamma} reaches its maximum, with
$p_{6,\max}=1/8$ and $p_{8,\max}=1/4$.
The relation to the order-two non-local magic is especially direct.  From
Eq.~\eqref{eq:magicgamma},
$u^2(1-u^2)=1-2^{-\mathcal M_2^{\rm Sch}}$, and hence
\begin{equation}
 p_6=\frac12\left(1-2^{-\mathcal M_2^{\rm Sch}}\right),
 \qquad
 p_8=1-2^{-\mathcal M_2^{\rm Sch}}.
 \label{eq:MagicConcentrationProbabilities}
\end{equation}
Equation~\eqref{eq:MagicConcentrationProbabilities} relates the input
state's non-local stabilizer R\'enyi entropy to the success probabilities
of the specified protocol after global Clifford decoding.  Independent
repetition of the eight-copy block gives the achievable asymptotic rate
\begin{equation}
 R(\phi_\gamma\rightarrow\mathrm{CCZ})
 \geq\frac{M_3^{\rm lin}(\phi_\gamma)}{12}
 =\frac18u^2(1-u^2).
 \label{eq:CCZAsymptoticRate}
\end{equation}
The concentration protocol uses independently prepared copies, the
cross-partition CNOT, Clifford gates, Pauli measurements, feed-forward,
and postselection.  For spatially separated subsystems, implementing that
CNOT requires resources beyond LOCC.  The protocol follows
Ref.~\cite{Rizzo2026Concentration}.

\subsection{Symmetry resolution and the static resource dictionary}

\paragraph{Local-to-global parity reduction.}

In the symmetry-coupled setting, the conserved particle number can be replaced
here by a discrete parity to obtain a useful structural comparison
\cite{Himmelsbach2026Symmetry}.  Define
\begin{equation}
 \sigma_z^A=\sigma_z\otimes I_B,
 \qquad
 \sigma_z^B=I_A\otimes \sigma_z,
 \qquad
 P=\sigma_z^A \sigma_z^B=\sigma_z\otimes \sigma_z.
\label{eq:LocalAndGlobalParityOperators}
\end{equation}
In the computational basis, $P\vert{}00\rangle = \vert{}00\rangle$, $P\vert{}01\rangle = -\vert{}01\rangle$, $P\vert{}10\rangle =-\vert{}10\rangle$, $P\vert{}11\rangle = \vert{}11\rangle$. Thus $P = \operatorname{diag}(+1, -1, -1, +1)$.
Because $H_0$ in Eq.~\eqref{eq:twoqubit} is diagonal,
\begin{equation}
 [H_0,\sigma_z^A]=[H_0,\sigma_z^B]=[H_0,P]=0.
 \label{eq:H0LocalParitySymmetries}
\end{equation}
By contrast, the generator $G=\sigma_x\otimes \sigma_x$ anticommutes with either local
parity but commutes with their product,
\begin{equation}
 \{G,Z_A\}=\{G,Z_B\}=0,
 \qquad
 [G,P]=0,
 \label{eq:GeneratorParityAlgebra}
\end{equation}
since $\{\sigma_x, \sigma_z\} = 0$ and $G P = (\sigma_x \otimes \sigma_x)(\sigma_z \otimes \sigma_z) = (\sigma_x \sigma_z) \otimes (\sigma_x \sigma_z) = (-i\sigma_y) \otimes (-i\sigma_y) = -\sigma_y \otimes \sigma_y=PG$.
It follows that $[S_\gamma, P] = [e^{\gamma G}, P] = 0, \quad [S_\gamma^{-1}, P] = [e^{-\gamma G}, P] = 0$
and hence
\begin{equation}
 [H_\gamma,P]=0,
 \qquad
 [H_\gamma,Z_A]\neq0,
 \qquad
 [H_\gamma,Z_B]\neq0
 \quad(\gamma\neq0),
 \label{eq:HgammaGlobalParityOnly}
\end{equation}
for the distinct eigenvalues assumed in the two-qubit realization.  Thus the
independent local $\mathbb Z_2$ labels of the reference Hamiltonian are reduced
to a single conserved global parity after the non-unitary similarity
deformation.  In this precise symmetry-classification sense, the model is a
minimal discrete analogue of the local-charge-to-global-charge reduction in
symmetry-coupled bipartite systems.

The comparison is not an identification with Bose--Hubbard tunneling.  In
raising and lowering operators,
\begin{align}
 X_A X_B
 &=\sigma_x^A\sigma_x^B
 =(\sigma_A^++\sigma_A^-)(\sigma_B^++\sigma_B^-)\notag\\
 &=\sigma_A^+\sigma_B^-+\sigma_A^-\sigma_B^+
  +\sigma_A^+\sigma_B^++\sigma_A^-\sigma_B^-\notag\\
 &=|10\rangle\langle01|+|01\rangle\langle10|
  +|11\rangle\langle00|+|00\rangle\langle11|.
 \label{eq:XXExchangeAndPairTerms}
\end{align}
The first two terms exchange an excitation, whereas the last two create or
annihilate a pair.  Consequently $X_A X_B$ conserves total parity but not total
particle number, and $S_\gamma=e^{\gamma X_A X_B}$ is a non-unitary similarity
map rather than a unitary hopping evolution.  This is therefore a
symmetry-level analogy; large-system chaotic crossover and continuous-number
conservation results are not transferred to the present model
\cite{Himmelsbach2026Symmetry}.

\paragraph{Parity-resolved entropy.}

If $\ket{0}$ and $\ket{1}$ are interpreted as empty and occupied fermionic
modes and global even parity is imposed as a superselection rule, the two
local parity sectors have weights $a_\gamma^2$ and $b_\gamma^2$.  The
symmetry-resolved linear-entropy decomposition then gives
\begin{equation}
 S_{\rm numb}
 =1-a_\gamma^4-b_\gamma^4
 =\frac{C_\gamma^2}{2},
 \qquad
 S_{\rm conf}=0,
 \qquad
 1-\Tr[(\rho_A^R)^2]=S_{\rm numb}+S_{\rm conf}.
 \label{eq:ParityResolvedLinearEntropy}
\end{equation}
This is the discrete-parity specialization of the number-resolved
decomposition \cite{Himmelsbach2026Symmetry}.
Each fixed local-parity block is one-dimensional, so all of the linear entropy
comes from uncertainty between parity sectors.  This statement is conditional
on the fermionic superselection rule and must not be read as vanishing
ordinary qubit entanglement.  Moreover, the random-matrix entanglement
transition assumes large chaotic subsystems and symmetry-breaking
inter-subsystem couplings and is not invoked for the present
four-dimensional isospectral family \cite{Himmelsbach2026Symmetry}.

The one-dimensionality of both local parity blocks also sharpens the relation
among diagnostics already used in this work.  Combining
Eqs.~\eqref{eq:IPRexplicit}, \eqref{eq:negativity}, and
\eqref{eq:ParityResolvedLinearEntropy} gives
\begin{equation}
 S_{\rm numb}
 =1-I_2^{\rm comp}
 =\frac12\tanh^2(2\abs\gamma)
 =2\mathcal N_{\rm PPT}^2,
 \qquad
 S_{\rm conf}=0.
 \label{eq:ParityIPRNegativityDictionary}
\end{equation}
Thus the linear entanglement, loss of computational-basis localization, and
PPT negativity are not independent on this minimal Schmidt manifold.  The
corresponding von Neumann decomposition has the same interpretation:
\begin{equation}
 S_A^{(\mathrm{vN},R)}
 =-a_\gamma^2\ln a_\gamma^2-b_\gamma^2\ln b_\gamma^2,
 \qquad
 \sum_{q_A}p_{q_A}S(\rho_{q_A})=0.
 \label{eq:ParityResolvedVonNeumannEntropy}
\end{equation}
All of the von Neumann entropy is therefore the Shannon uncertainty of the
two local $\mathbb Z_2$ sector weights.  This is an exact finite-dimensional
identity, whereas the near dominance of number entropy in weakly coupled
chaotic systems is a dynamical large-system statement
\cite{Himmelsbach2026Symmetry}.

\paragraph{Scale-resolved information and resource locking.}

The same von Neumann entropy has a compact information-lattice interpretation
without introducing an additional state parameter
\cite{Thakur2026LocalInformation}.
With logarithms in base two, the information retained by either one-qubit
subsystem is its entropy deficit from the maximally mixed state, while the
information supported only on the two-qubit scale is the mutual information.
For the present globally pure state this gives
\begin{equation}
 I_A^{(0)}=1-\frac{S_A^{(\mathrm{vN},R)}}{\ln2},
 \qquad
 I_B^{(0)}=1-\frac{S_A^{(\mathrm{vN},R)}}{\ln2},
 \qquad
 I_{AB}^{(1)}=\frac{2S_A^{(\mathrm{vN},R)}}{\ln2}.
 \label{eq:TwoQubitInformationLattice}
\end{equation}
Their sum is identically two bits.  At $\gamma=0$ the information is entirely
local, whereas in the Bell limit the two local contributions vanish and both
bits reside at the bipartite scale.  This finite two-site identity is exact;
the ballistic front, Tracy--Widom profile, and $t^{1/3}$ broadening of long
random circuits do not follow for the static two-qubit Hamiltonian considered
here \cite{Thakur2026LocalInformation}.

Resource equivalence under generic ergodic evolution arises dynamically only
after ensemble averaging \cite{Aditya2026ResourceEquivalence}.  The present
family is not ergodic, but its one-parameter Schmidt structure enforces an
exact but model-constrained kinematic relation among several state
functionals.  Let $\Delta_Z$ denote complete dephasing in the
global computational basis.  Since
\begin{equation}
 \Delta_Z(\rho_\gamma^R)
 =a_\gamma^2|00\rangle\langle00|
 +b_\gamma^2|11\rangle\langle11|,
 \qquad
 \Tr[\Delta_Z(\rho_\gamma^R)^2]
 =I_2^{\rm comp},
 \label{eq:GlobalComputationalDephasingPurity}
\end{equation}
the base-two R\'enyi-2 coherence is exactly
\begin{align}
 C_d(\rho_\gamma^R)
 &:=S_{2,\mathrm{bits}}(\Delta_Z[\rho_\gamma^R])
 -S_{2,\mathrm{bits}}(\rho_\gamma^R)\notag\\
 &=-\log_2 I_2^{\rm comp}
 =\frac{S_A^{(2,R)}}{\ln2}.
 \label{eq:CoherenceEntanglementLocking}
\end{align}
This equality holds because the computational basis is the Schmidt basis and
the global right state is pure; it is not a basis-independent identity between
coherence and entanglement.

The same reduced purity also fixes the exact order-two non-local magic.  Using
$u^2=2I_2^{\rm comp}-1$ in Eq.~\eqref{eq:magicgamma} gives
\begin{equation}
 \mathcal M_2^{\rm Sch}
 =-\log_2\left[
 4(I_2^{\rm comp})^2-6I_2^{\rm comp}+3
 \right].
 \label{eq:MagicPurityLocking}
\end{equation}
Equations~\eqref{eq:ParityIPRNegativityDictionary},
\eqref{eq:CoherenceEntanglementLocking}, and
\eqref{eq:MagicPurityLocking} form an exact static resource dictionary.  Its
origin is restriction to a one-dimensional curve in state space, not a
low-dimensional dynamical manifold produced by scrambling
\cite{Aditya2026ResourceEquivalence}.  In particular,
$\rho_\gamma^R=(\rho_\gamma^R)^T$ is real in the computational basis, so its
imaginarity vanishes for every $\gamma$ even though $C_d$ is generally
nonzero.  The Haar-ergodic coherence--imaginarity relation of that reference
therefore cannot be transferred to this family.  Likewise, its two-replica
theory does not by itself derive the higher-replica nonstabilizerness in
Eq.~\eqref{eq:magicgamma}; Eq.~\eqref{eq:MagicPurityLocking} follows instead
from the exact Schmidt spectrum.

\endgroup

\section{Discussion and conclusion}
\begingroup
The similarity deformation preserves the Jordan structure of $H_0$ at
all finite $\gamma$ while changing right eigenvectors relative to the
fixed Hilbert-space inner product and bipartition.  The transported
canonical vectors evolve from product states toward Bell states, while
the degenerate right eigenspace also contains separable vectors.  The
basis-independent maximum over that eigenspace behaves differently from
the magic of an individual canonical eigenvector.  In a quasi-Hermitian
representation, the metric and observables transform together; the
resource analysis here instead uses the fixed-inner-product convention.

The right-state, biorthogonal, and singular-value reductions
probe different properties.  The positive right-state reduction gives the
ordinary von Neumann and R\'enyi entanglement entropies.  The eigenvalue-based
biorthogonal reduction diagnoses non-positivity and may be negative or
complex.  Its ordinary SVD entropy reconstructs the right-state entropy for
the canonical diagonal pair, while the BUISVD entropy is real, bounded, and
invariant under admissible biorthogonal rescalings but reduces here to a
rank-sensitive quantity.  These diagnostics are complementary rather than
interchangeable.

Page entropy and Haar-averaged purity provide complementary typicality
benchmarks for the complete right eigenspace.  Their use must respect whether
the sampled states are complex or real, whether the relevant sector
factorizes as $\cH_A\otimes\cH_B$, and whether an exact symmetry such as fixed
particle number restricts the ensemble.  The projected sampling prescription
therefore supplies the appropriate basis-independent numerical benchmark
without assigning physical meaning to an arbitrary basis choice inside the
degenerate eigenspace.

The stabilizer analysis adds a distinction that entropy cannot make.  For the
present rank-two state, the Schmidt representative is globally optimal over
unrestricted local unitaries, so
\begin{equation}
 \mathcal M_2^{\rm NL}(\psi_\gamma)
 =\mathcal M_2^{\rm Sch}(\psi_\gamma)
 =-\log_2(1-u_\gamma^2+u_\gamma^4).
 \label{eq:ConclusionExactNonlocalMagic}
\end{equation}
For the canonical transported eigenvectors this exact non-local resource
vanishes at both the product and Bell endpoints and peaks at an intermediate
nonuniform Schmidt spectrum, whereas
$S_A^{(\mathrm{vN},R)}$, $S_A^{(2,R)}$, and $\mathcal N_{\rm PPT}$ grow
monotonically.  In contrast, optimizing over the entire degenerate right
eigenspace yields a minimum of zero and a maximum that equals the canonical
magic only below $|\gamma|=\gamma_*$, saturating thereafter at
$\log_2(4/3)$ even when the canonical magic tends to zero,
Eq.~\eqref{eq:SubspaceMaximumMagic}.  These extrema are invariant under
changes of basis within the eigenspace.  Near the product endpoint the
canonical-state resource has quadratic onset,
$\mathcal M_2^{\rm NL}=C_\gamma^2/\ln2+O(C_\gamma^4)$, as shown in
Eq.~\eqref{eq:WeakDeformationMagicScaling}.  For $m$ identical imperfect
dimers, collective local-unitary
optimization further changes the apparent extensive fixed-basis SRE into
\begin{equation}
 \mathcal M_2^{\rm NL}(|\psi_\gamma\rangle^{\otimes m})
 =\frac32\log_2m+O_\gamma(1),
 \qquad 0<|\gamma|<\infty,
 \label{eq:ConclusionMulticopyScaling}
\end{equation}
which makes the distinction between pairwise processing and true non-local
resource optimization explicit.

The operational extensions are different realizations of this same
one-parameter Schmidt geometry rather than independent state variables.  The
Triangle witness and Schmidt--MUB tests diagnose noisy magic and target-state
fidelity, while the correlation tensor supplies an exact CHSH diagnostic.
Local dephasing leaves
the one-body entropy unchanged while suppressing negativity and CHSH
violation, demonstrating that subsystem entropy ceases to be an entanglement
measure once the global state is mixed.  The rotation--CNOT circuit and the CCZ concentration protocol supply
global-Clifford operations for the decoded state, with cross-partition
processing as specified in Sec. VII.  The
parity-resolved decomposition and static resource dictionary express
these consequences through the common coordinates
$u_\gamma=\operatorname{sech}(2|\gamma|)$ and
$C_\gamma=\sqrt{1-u_\gamma^2}$.

In summary, the complementary diagnostic quantities used in the two-qubit
and larger-dimensional realizations can be organized schematically as
\begin{equation}
 \left(r,\;\kappa_2,\;\Delta_{\mathrm{vN}},\;\Delta_2^{\mathrm{ann}},\;
 \mathcal N_{\rm PPT},\;\mathcal M_2^{\rm NL},\;
 \mathcal M_{2,\max}^{\mathcal E_0^R}\right),
 \label{eq:ConclusionDiagnosticTuple}
\end{equation}
These seven entries respectively describe eigenspace multiplicity,
the conditioning of the similarity map, the von Neumann and annealed
R\'enyi-2 deficits from full-space Haar benchmarks, PPT negativity,
the non-local magic of a selected canonical right eigenvector, and the
basis-independent maximum of non-local magic over the complete degenerate
right eigenspace.  They form a collection of complementary diagnostics,
not seven jointly defined observables of a single state: the Haar deficits
are illustrated in the larger-dimensional example, while PPT negativity
and the two magic quantities are evaluated exactly in the two-qubit model.
The framework therefore distinguishes algebraic degeneracy, numerical
conditioning, restricted entanglement statistics, and stabilizer resources,
while specifying the physical inner product, tensor-product structure,
symmetry sector, and whether a statement concerns a chosen eigenvector
or the entire degenerate eigenspace.
\endgroup


\appendix
\section{Entropy--concurrence dictionary for dissipative stabilization}
\label{app:EntropyConcurrence}

\begingroup

The correspondence also turns the static entropies into a dynamical resource
parameter.  The concurrence of Eq.~\eqref{eq:dissipativeSchmidtState} is
\begin{align}
 C_\gamma
 &=2a_\gamma b_\gamma
 =\frac{2\cosh(\abs{\gamma})\sinh(\abs{\gamma})}
 {\cosh(2\abs{\gamma})}
 =\tanh(2\abs{\gamma})
 =w_\gamma.
 \label{eq:dissipativeConcurrence}
\end{align}
Since $p_+p_-=a_\gamma^2b_\gamma^2=C_\gamma^2/4$, the reduced purity becomes
\begin{align}
 \Tr[(\rho_A^R)^2]
 &=(p_++p_-)^2-2p_+p_-
 =1-\frac{C_\gamma^2}{2}.
 \label{eq:purityConcurrenceRelation}
\end{align}
Combining this identity with Eq.~\eqref{eq:exactS2} gives the two equivalent
relations
\begin{equation}
 S_A^{(2,R)}= -\ln \text{Tr}[(\rho_A^R)^2]=-\ln\left(1-\frac{C_\gamma^2}{2}\right),
 \qquad
 C_\gamma^2=2\left(1-e^{-S_A^{(2,R)}}\right).
 \label{eq:entropyConcurrenceDictionary}
\end{equation}
Thus the second R\'enyi entropy fixes the strength of the entangled resource
entering the reservoir protocol.

\endgroup

\end{document}